\PassOptionsToPackage{hyphens}{url}
\documentclass[conference]{IEEEtran}

\usepackage{tikz}
\usepackage{amsmath}
\usepackage{filecontents}
\usepackage[utf8]{inputenc}
\usepackage{comment}
\usepackage{graphicx}
\graphicspath{ {./images/} }
\usepackage[numbers,sort]{natbib} 

\usepackage{tabularx}
\usepackage{enumitem}
\usepackage[export]{adjustbox}
\usepackage{xcolor}
\usepackage{textgreek}
\usepackage{multirow}
\usepackage{bigstrut}
\usepackage{xspace}
\usepackage[caption=false]{subfig}
\usepackage{textcomp}
\usepackage{hyperref}
\hypersetup{
    colorlinks,
    linkcolor={red!50!black},
    citecolor={green!60!black},
urlcolor={blue!80!black}}

\usepackage{amsmath,amssymb,amsfonts}
\usepackage{algorithmicx}
\usepackage{graphicx}
\usepackage{textcomp}
\usepackage{xcolor}
\usepackage{booktabs}
\usepackage[capitalise,nameinlink,noabbrev]{cleveref}
\def\BibTeX{{\rm B\kern-.05em{\sc i\kern-.025em b}\kern-.08em
    T\kern-.1667em\lower.7ex\hbox{E}\kern-.125emX}}
\usepackage{svg}
\usepackage{bm}
\usepackage{fancyhdr}
\usepackage{algorithm}
\usepackage{algpseudocode}
\usepackage{tikz}
\definecolor{darkbrown}{rgb}{0.55, 0.39, 0.24}
\newcommand*\circleddark[1]{\tikz[baseline=(char.base)]{
            \node[shape=circle,draw=darkbrown,inner sep=1pt,fill=darkbrown] (char) {\textbf{\textcolor{white}{#1}}};}}
\newcommand*\circled[1]{\tikz[baseline=(char.base)]{
            \node[shape=circle,draw=darkbrown, thick,inner sep=1pt] (char) {\textbf{#1}};}}
\usepackage{xcolor,pifont}
\newcommand*\colourcheck[1]{
  \expandafter\newcommand\csname #1check\endcsname{\textcolor{#1}{\ding{52}}}
}
\colourcheck{blue}
\colourcheck{green}
\colourcheck{red}

\newcommand{\parhead}[1]{\vspace{3pt plus 1pt minus 3pt}\par\noindent\textbf{#1}\hspace{.4em plus .2em minus .2em}}

\newcommand{\email}[1]{{\rm\textsf{\href{mailto:#1}{#1}}}}

\newcommand{\bt}{Borrowed Time\xspace}
\makeatletter
\newcommand{\linebreakand}{
  \end{@IEEEauthorhalign}
  \hfill\mbox{}\par
  \mbox{}\hfill\begin{@IEEEauthorhalign}
}
\makeatother
\usepackage{hyperref}
\begin{document}
\title{On Borrowed Time -- \\Preventing Static Side-Channel Analysis}

\author{\IEEEauthorblockN{Robert Dumitru} 
\IEEEauthorblockA{\textit{Ruhr University Bochum, Germany \&}\\
\textit{\phantom{fffffffff} The University of Adelaide, Australia \phantom{fffffffff}}\\
\small\email{robert.dumitru@adelaide.edu.au}}
\and
\IEEEauthorblockN{Thorben Moos}
\IEEEauthorblockA{\textit{\phantom{fffffffffffffffffff} UC Louvain, Belgium \phantom{ffffffffffffffff}}\\
\small\email{thorben.moos@uclouvain.be}}
\linebreakand
\IEEEauthorblockN{Andrew Wabnitz}
\IEEEauthorblockA{\textit{Defence Science and Technology Group, Australia}\\
\small\email{andrew.wabnitz1@defence.gov.au}}
\and
\IEEEauthorblockN{Yuval Yarom}
\IEEEauthorblockA{\textit{\phantom{ffffff} Ruhr University Bochum, Germany \phantom{ffffffff}}\\
\small\email{yuval.yarom@rub.de}}
}

\IEEEoverridecommandlockouts
\makeatletter\def\@IEEEpubidpullup{6.5\baselineskip}\makeatother 
\IEEEpubid{\parbox{\columnwidth}{
        Network and Distributed System Security (NDSS) Symposium 2025\\
        24-28 February 2025, San Diego, CA, USA\\
        ISBN 979-8-9894372-8-3\\
        https://dx.doi.org/10.14722/ndss.2025.241243\\
        www.ndss-symposium.org
}
\hspace{\columnsep}\makebox[\columnwidth]{}}

\maketitle
\bstctlcite{IEEEexample:BSTcontrol}

\begin{abstract}

In recent years a new class of side-channel attacks has emerged.
Instead of targeting device emissions during dynamic computation, adversaries now frequently exploit the leakage or response behaviour 
of integrated circuits in a static state.
Members of this class include Static Power Side-Channel Analysis (SCA), Laser Logic State Imaging (LLSI) and Impedance Analysis (IA).
Despite relying on different physical phenomena, they all enable the extraction of sensitive information from circuits in a static state with high accuracy and low noise -- a trait that poses a significant threat to many established side-channel countermeasures.

In this work, we point out the shortcomings of existing solutions and derive a simple yet effective countermeasure.
We observe that in order to realise their full potential, static side-channel attacks require the targeted data to remain unchanged for a certain amount of time.
For some cryptographic secrets this happens naturally, for others it requires stopping the target circuit's clock.
Our proposal, called \bt, hinders an attacker's ability to leverage such idle conditions, even if full control over the global clock signal is obtained.
For that, by design, key-dependent data may only be present in unprotected temporary storage (e.g.\ flip-flops) when strictly needed.
\bt then continuously monitors the target circuit and upon detecting an idle state, securely wipes sensitive contents. 

We demonstrate the need for our countermeasure and its effectiveness by mounting practical static power SCA attacks against cryptographic systems on FPGAs, with and without \bt.
In one case we attack a masked implementation and show that it is only protected with our countermeasure in place. 
Furthermore we demonstrate that secure on-demand wiping of sensitive data works as intended, affirming the theory that the technique also effectively hinders LLSI and IA.

\end{abstract}

\section{Introduction}
The seminal work of \citet{Kocher96} demonstrated that implementations of mathematically secure cryptographic primitives can be vulnerable to attacks via side-channel analysis (SCA) exploiting the leakage of sensitive information through physical properties of the implementation.
Since then, multiple side-channels have been demonstrated
exploiting various effects, such as timing~\cite{Bernstein05, BB03}, 
power consumption~\cite{KocherJJ99, MangardOP07},
electromagnetic emanations~\cite{QS01, GMO01},
shared micro-architectural components~\cite{GeYCH18, Tromer2010}, even acoustic~\cite{GenkinST14} and photonic emanations~\cite{Schlosser12, KN+13}.
Among the physical sources of side-channel leakage, exploitation of power consumption, known as power analysis, has received the most attention. 

Research on power analysis has historically focused on attacks exploiting the instantaneous power consumed during computation, also known as dynamic power analysis.
However, with the continued down-scaling of Complementary Metal-Oxide-Semiconductor (CMOS) technology, the relative weight of static power---used for maintaining the logical state of a circuit---has increased as part of the total power budget~\cite{Narendra01, Xue10, Kao02} and so too has the relative exploitable static power side-channel leakage. 
This has shifted recent attention towards static power SCA~\cite{Moradi14, DelPozo15, Moos17, Bellizia17, Moos20, Bellizia21, Karimi2019, Moos_2020, Moos_2019}.

\begin{figure}[t]
    \centering
    \includegraphics[width=0.9\linewidth]{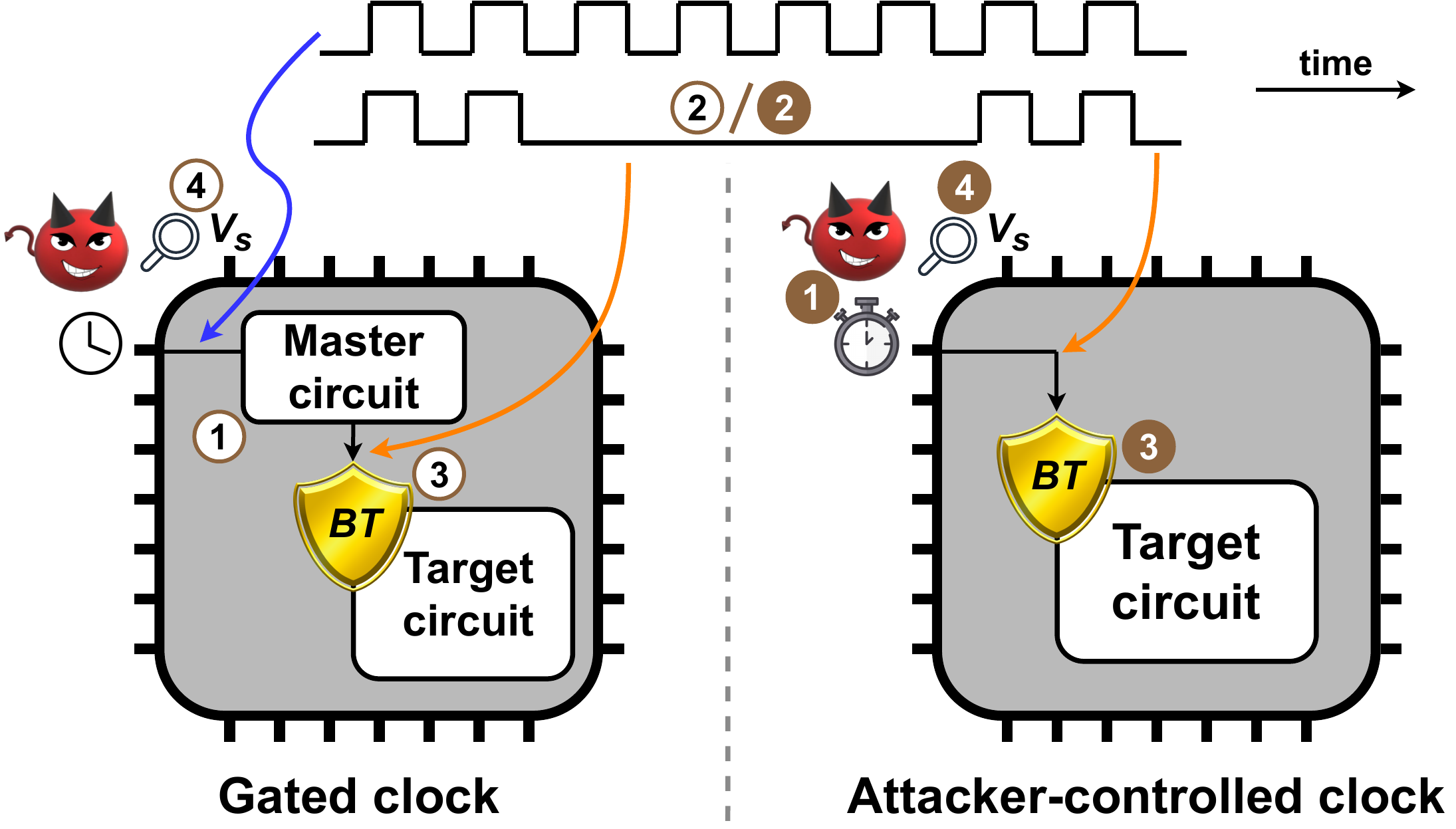}
    \caption{Static power SCA attack stages in the face of \bt countermeasure: (1) (2) -- The attacker leverages a stopped clock for prolonged static measurement period. (3) -- \bt detects clock stoppage and clears sensitive data. (4) -- The attacker can no longer observe static leakage.}
    \label{fig:topdiagram}
\end{figure}

In a typical static power SCA attack, the adversary exploits a prolonged state of the target circuit -- see \cref{fig:topdiagram} \circled{2} / \circleddark{2} -- when it is known to contain stable secret data in some of its state registers. 
This prolonged state is induced either by a master circuit that dynamically disables the clock provided to the target circuit \circled{1} (e.g.\ via clock gating to reduce power consumption~\cite{ClockGating}), or directly by the attacker \circleddark{1} if they can stop the clock signal. 
The attacker waits for some time to allow the dynamic effects of prior computation to subside before accurately measuring the global leakage current of the device.
The latter can be achieved by dedicated equipment for measuring nano-scale currents~\cite{Bellizia17, Moos21}, or by averaging a large number of power measurements taken using an oscilloscope \circled{4}/\circleddark{4} over the prolonged period, resulting in high measurement accuracy and noise reduction~\cite{Moos20,Moos_2019}.

Beyond static power SCA attacks,
there exist data extraction methods with potentially even higher accuracy, which similarly exploit the vulnerability of circuits in a static state.
Such attacks include Laser Logic State Imaging (LLSI)~\cite{Krachenfels21} and Impedance Analysis (IA)~\cite{Monfared23}. 
The former is a type of optical probing based on the reflection of laser light by transistors under modulated supply voltage, while the latter exploits variations in the impedance of the measured system at different frequencies, which correlate with data temporarily stored on a chip.
Both techniques have the decisive advantage over measurements of the global static power of a chip in that they can better localise specific state registers and target their data values.
From hereon we refer to the grouping of these methods with static power SCA as static SCA attacks.
Crucially, all static SCA attacks can be performed in many cases even when the adversary does not have outright clock control.
This is for example the case when the target device keeps secret information in unprotected temporary storage like flip-flops for an extended period of time under normal operation~\cite{Moos_2019, Monfared23} -- a condition that is not uncommon and might even be induced by additional power-optimising clock-gating logic that is introduced by Electronic Design Automation (EDA) tools during a system build, unbeknownst to the designer~\cite{vivug}.

Since the first practical demonstration of static SCA attacks~\cite{Moradi14}, research in the area has demonstrated attacks on various cryptographic primitives~\cite{Moos17, Bellizia17}, conducted investigations of attack factors~\cite{Moos20}, and attack improvements~\cite{Bellizia17, Moos17, Moos_2019, Moos20}.
In particular, past work has demonstrated static SCA to be effective even against targets that implement countermeasures which hinder dynamic power analysis~\cite{Moos_2019, Moos17, Bellizia16, Bellizia17uni, Moos21}, demonstrating viable attacks against securely masked cryptographic primitives~\cite{Moos17, Moos_2019, Moos21, Moradi14, Monfared23}.
However, despite the clear need for dedicated protection solutions, little research has gone towards designing countermeasures against the entire class of static SCA.
In this work we set out to fill this gap.

\subsection*{Our Contribution}
We first observe that all published static SCA attacks require a relatively long period, spanning at least several hundreds of typical clock cycles (considering MHz frequencies), in which two conditions hold simultaneously:
\begin{enumerate}[nosep,leftmargin=*]
\item The clock signal supplied to the target circuit is stopped. 
\item Registers contain sensitive data  and thereby expose it to leakage.
\end{enumerate}

The core idea behind our countermeasure, called \bt, is to prevent situations in which both conditions hold.
Specifically, \bt continuously monitors the incoming clock signal. 
If \bt detects that the clock is stopped for too long, it performs a masked clear, wiping sensitive register contents by overwriting them with random values, to prevent introducing dynamic leakage through the action of wiping. 
\bt is deployed in the same fabric as the target circuit (\cref{fig:topdiagram} \circled{3}/\circleddark{3}).

We propose two approaches for implementing the clock-monitor in \bt.
One uses typical clock management circuitry found in conventional digital systems and imposes minimal design-effort overhead; the other uses a custom asynchronous circuit which is better suited to low-power designs which may also employ clock-gating.

The first approach uses a Phase-Locked Loop (PLL), which is a component commonly used for clock management in digital systems. 
In a nutshell, a PLL uses a feedback loop to ensure that an output clock (or clocks), which it distributes across the circuit, remains synchronised with an input reference clock that is typically provided by an external source such as a crystal oscillator or another circuit.
Stopping the input clock breaks synchronisation between it and the output clock, allowing the PLL to rapidly detect such an incident.

While PLLs are effective at detecting a stopped clock, they have two main limitations. 
First, PLLs are relatively large and may therefore be too expensive in terms of power and area consumption for some uses. 
Second, because PLLs take a long time to startup by synchronising the output and input clocks, they are unsuitable in some deployment scenarios.
One such example is within clock-gated circuits, where a master circuit (e.g.\ \cref{fig:topdiagram} \circled{1}) dynamically disables the clock signal provided to some circuit components in order to reduce power consumption when these components are not active~\cite{ClockGating}. 

To accommodate for cases where PLLs may not be suitable, we propose an alternative design based on an asynchronous system that samples the clock signal at multiple points in time and monitors its natural variation.
Where this variation is absent and the clock value is the same across all sampled points in time, the design indicates a stopped clock.
This design variant also facilitates the secure implementation of clock-gating within a system while imposing minimal overhead, particularly when this overhead is offset by the substantial energy savings brought about by clock-gating.

To test \bt's effectiveness, we perform practical static power SCA attacks against cryptographic circuits with and without the countermeasure in place.
We implement an unprotected \texttt{AES} circuit and a masked \texttt{SKINNY} circuit on different Field-Programmable Gate Array (FPGA) platforms, and we perform end-to-end static power SCA attacks against them to recover their secret keys.
With our setups, we observe information leakage through static power in both implementations.
The unprotected \texttt{AES} implementation exhibits strong leakage that benefits the attacker.
We recover the full \texttt{AES} key using 1,500 samples, each sample being a measured power trace from a single encryption.
The \texttt{SKINNY} implementation is also vulnerable despite its masking protection, requiring 16,000 samples for a subkey recovery.

Detecting whether a clock has stopped inevitably takes some time.
At minimum, the detection mechanism needs to wait and see a missed clock edge.
For \bt to be an effective countermeasure, detection and clear time must be shorter than the time needed to carry out a static SCA attack, which is typically in the range of milliseconds.
Optical attacks like LLSI are reported to require multiple milliseconds per pixel~\cite{Krachenfels21}.
The Vector Network Analyser (VNA) used for IA lists the time needed for a frequency sweep in milliseconds~\cite{Monfared23} (concrete values depend on various parameters), though the authors rightfully point out that adversarial knowledge of the frequencies to target can speed up the attack significantly.
In static power SCA attacks,  measurement accuracy is notably reduced by increasing the acquisition speed (i.e.\ shortening and shifting the acquisition time window closer to the last clock pulse), due to 1) reduced opportunity for noise reduction through averaging~\cite{Moos20} and 2) the \emph{memory effect} --
a period where dynamic effects from prior computation affect the power consumption measured.
While some past works report the wait time (from when the clock is stopped until measurements start to be taken) used for attacks to be in the order of milliseconds~\cite{Moos17, Moos20}, little information is provided in support of the choice of this delay.
In general, it is clear from the literature that all reported static SCA attacks suffer from increased noise and decreased accuracy when the time window for the attack shrinks.

Because \bt relies in part on static leakage not being instantly exploitable once the clock is stopped, we evaluate how much wait time is needed to mount a static power SCA attack.
In our experiments, the attack becomes infeasible with wait times any shorter than 200\,\textmu s, whereas \bt can eliminate leakage in less than 1\,\textmu s, assuming the target hardware operates in at least the MHz range. 

To test \bt's effectiveness, we incorporate it in each cryptographic target.
We demonstrate that both implementations 
prevent the attacks, and the protected targets do not show evidence of leakage even with 1 million samples.

To summarise, we make the following contributions:
\begin{itemize}[nosep,leftmargin=*,topsep=-12pt]
\item We present two designs of \bt, a countermeasure for static SCA attacks that detects a stopped clock and performs a masked clear on the sensitive data within as little as one clock cycle, well before the time window in which an attacker has the opportunity to mount an attack (\cref{section:countermeasure}).
\item We investigate the memory effect in the context of static power SCA, demonstrating that an attacker needs to wait at minimum several hundreds of clock cycles worth of time after stopping the clock before they can observe static leakage that would be usable for mounting an attack (\cref{section:memory}).
\item We practically implement \bt within \texttt{AES} and masked \texttt{SKINNY} implementations on different FPGA technologies and evaluate them by mounting end-to-end static power SCA-based key recovery attacks with varying setups, demonstrating that \bt effectively mitigates these attacks, even where masking does not (\cref{section:evaluation}). 
\end{itemize}

\section{Background}
\label{section:backg}

\subsection{Physical Side-Channel Analysis}

The physical interactions of electronic devices with their environment can leak information about the computations they perform and values processed.
This leakage manifests through inherent correlations between the computation or data being processed with the observed physical effects. 
The practice of exploiting such unintended sources of information leakage to reveal secrets is known as physical side-channel analysis. 

Since the seminal work of \citet{Kocher96}, SCA has become a significant threat to cryptographic implementations in particular, 
because visibility of computational intermediates defies the black-box assumption\footnote{Black-box: adversaries can only observe inputs and corresponding outputs.} that upholds the security of cryptographic primitives.
Side-channel attacks are categorised by the effect responsible for leaking information, with significant effort being invested in power analysis, which exploits the power consumption of a target device~\cite{MangardOP07, KocherJJ99}. 
Although physical access to the target device is typically required, remote power analysis attacks through cloud-integrated FPGAs~\cite{Zhao18} and software~\cite{Mantel18, Lipp21} have also recently been demonstrated.

\subsection{Static Side-Channel Analysis}
Static SCA attacks are physical hardware attacks which exploit the leakage or response behaviour of integrated circuits (ICs) when in a stable, static state.
We further divide this family into power (which our experiments are based on) and non-power static SCA attacks.

\subsubsection{Static Power SCA Attacks}
The majority of power analysis attacks historically explored is concerned with the dynamic power consumption of target devices, i.e.\ the instantaneous consumption associated with changing the logical state of electronic gates.
In contrast, static power SCA is concerned with the power dissipated to maintain the logical state of the device, which exhibits a dependence on the stored data~\cite{Giorgetti07, ABBAS2014179}.

Any digital circuit can be abstracted as a system made up of pipelines of combinatorial logic elements sectioned between sequential (state) elements.
The core idea of exploiting static power information leakage is that a target device can be held in a certain state (represented by the contents of all state register elements) by stopping the clock signal fed to the device for a long enough time such that: all dynamic effects from transitioning into the given state have died off, and measurements can be taken across a sufficiently long window to average away a large amount of noise.
Static power SCA considers this averaged univariate random variable (the measurement \emph{trace}) as representative of a target's given state, whereas dynamic power SCA considers an instantaneous power trace as a time-dependent multivariate random variable representative of a computation performed by the target (related to its transition between states).
Static power SCA adversaries must be able to make use of clock manipulation, however, as~\citet{Moos_2019} shows this does not strictly require a stronger threat model (than dynamic) where the attacker outright controls the target's clock signal, instead they can leverage the interrupted clock signal provision to a clock-gated target (\cref{fig:topdiagram} left side). 

The earliest reports on this class of attack were based on simulated analysis~\cite{Giorgetti07, Lin08} and 
the first practical demonstration was performed by \citet{Moradi14} in 2014 against FPGA targets.
Since then, more works~\cite{DelPozo15, Moos17, Bellizia17, Moos20, Bellizia21, Karimi2019, Moos_2020, Moos_2019} have practically evaluated static power SCA attacks. 
A comprehensive history of the research area is described in~\cite{Moos20, Moos21}.

\parhead{Technology Down-Scaling.}
The contribution of static leakage to overall power consumption among modern CMOS technology is becoming proportionally larger~\cite{Narendra01, Xue10, Kao02} with the continued down-scaling of semiconductors as a proponent of Moore's law~\cite{Moore65}.
Simulations  focusing on 
nanometer CMOS technologies show a direct increase in static leakage with smaller scale technologies~\cite{Lin08}.
From practical results, \citet{Moradi14} demonstrated that the \emph{absolute} leakage current and leakage related to data contents does not necessarily directly increase with smaller technologies between different FPGA families, although this does not mean the same holds for relative proportions.
The tested technology families had more differences than just scale however, the details of which are not publicly available and could potentially influence leakage.
\citet{Moos_2019} performs similar testing with ASICs (Application-Specific Integrated Circuits), finding a direct increase in data-dependent static leakage across shrinking CMOS technologies.

\parhead{Measurement Factors.}
In static power SCA, while keeping a target circuit in a certain state via some form of clock control, many power measurements are taken over a long measurement window which are then averaged into a single \emph{trace} value representative of its given state.
This \emph{intra-trace averaging} reduces noise exponentially~\cite{Moos20}.
The most prominent physical effects responsible for quiescent current flow in CMOS circuits are sub-threshold leakage current and reverse biased PN junction current~\cite{Bellizia17uni}. 
\citet{Moos20} investigate the influence of voltage and temperature on this static leakage, finding an exponential dependence on operating temperature, while increases in operating voltage only marginally increase leakage.
\citet{Moos_2019} shows that the influence of temperature on static leakage is greater for smaller process technologies. 
Moreover, the nonlinear temperature dependencies can be leveraged 
to conduct multivariate attacks~\cite{Bellizia17tem, Djukanovic17}. 

\parhead{Memory Effect and Measurement Interval.}
To exploit leakage of static register contents an attacker must wait ample time for the dynamic effects within the target circuit to settle before taking measurements.
One might therefore expect it to be sufficient to wait until all of the signal value changes from a clock transition are propagated to the end of their paths.
However, it turns out that dynamic effects continue to be observable in the measurements of a circuit's power consumption profile well beyond this time, even extending across several clock cycles in a phenomena coined the \emph{memory effect} in~\cite{Moradi13}.
This may arise from slow transient response of the leakage current among individual CMOS elements, and/or from aggregated system effects like reflections, both inside the device under test and in interaction with the (active) components of the measurement setup. 
\citet{Moradi13} use the memory effect to combine leakages from operations across multiple clock cycles into univariate readings.
Against their FPGA target, this significantly extends the window of usable dynamic leakage by several microseconds. 

Past works observe that the memory effect is highly influenced by the power measurement setup used~\cite{Moradi13, Moos17, Moos20}.
These works use very similar setups; both incorporate a high-gain low-bandwidth DC amplifier. 
The works mention that from their analysis the memory effect influences 
static power measurement for 20\,ms, but go into no further detail on how this was evaluated.

\subsubsection{Non-Power Static SCA Attacks}
Laser Logic State Imaging~\cite{Krachenfels21} (LLSI) is a recently proposed laser-assisted SCA technique that can be used to directly observe the states of circuit registers during a certain stopped clock cycle.
By modulating the target's voltage supply and imaging its signal lines using lasers, they are able to observe different reflected responses dependent on the state (stored 1 or 0). 
Though this approach requires knowledge of sensitive register's physical locations, LLSI can also be used to assist their discovery~\cite{Krachenfels_21}.

Impedance Analysis (IA)~\cite{Monfared23} similarly involves modulating the circuit's supply voltage during a stopped clock cycle, but instead performing a frequency sweep and observing the system response.
This response varies based on the system's overall impedance which is dependent on circuit state, and thereby leaks information.
Moreover, registers at certain locations are responsible for the responses at certain frequencies.

Both of these techniques present unique methods of observing the static state of a circuit with different, higher-resolution representations and localisation features compared to static power SCA (while also requiring far more sophisticated resources to mount). 
Crucially however, as with static power SCA, both methods when performed as described by their authors require stopping the clock signal fed to the target circuit for a prolonged period.

\subsection{Countermeasures} 
Countermeasure approaches against physical SCA fall into two main categories: information masking, which algorithmically amplifies existing noise; and information hiding, which either decreases the exploitable signal (a.k.a. equalisation) or increases the noise (a.k.a. randomisation)~\cite{MangardOP07}.

Masking or `secret sharing' is a widely adopted class of countermeasures, in which the sensitive values used for computation within a system are never explicitly stored or processed in the system at any given moment~\cite{ChariJRR99}.
For example, a sensitive intermediate bit $x$ is encoded as a set of randomly generated shares $x_0$, $x_1$, ..., $x_n$, such that $x = x_0 \oplus x_1 \oplus ... \oplus x_n$.
The data is `split' and processed in the form of these shares and combined at the end to produce the output.
Masking schemes, especially those based on Boolean encodings, are known to work as intended only if the adversarial observations made are sufficiently noisy~\cite{Masure23}.
This condition, however, tends to be violated in presence of static SCA adversaries.
For example, previous works have indicated that attackers leveraging static power SCA can exploit higher-order leakages of masked systems with much lower data complexity than attackers leveraging dynamic leakage~\cite{Moos17, Moos_2019}.
With the low noise and precise localisation capabilities of non-power static SCA attacks like LLSI and IA, masking can sometimes even be completely circumvented~\cite{Krachenfels21, Monfared23}.  
The security of countermeasure schemes that rely on randomness is inextricably linked to the security of the source of randomness. 
\citet{Cassiers23} assess the costs of various methods of generating cryptographically-strong random bits for applications like masking, ultimately recommending a hybrid system for high-security settings that involves using a cryptographically secure Pseudo-Random Number Generator (e.g.\ Trivium~\cite{Trivium}) which takes an initial 80-bit seed from a True Random Number Generator (e.g.\ ES-TRNG~\cite{Yan18}) at power-up.

Logic balancing schemes~\cite{Tiri02, Tiri04, Popp05, Nawaz17, Fadaeinia21, Moos21} are a hiding class of countermeasures that aim to balance the data-dependence of power dissipation of logic elements. 
They use complementary signals to balance the number of bits toggled and stored in computation.
These logic styles rely on perfectly balanced power consumption from the underlying transistors and signal routing, which is very difficult to achieve in practice due to semiconductor process variation~\cite{Moos21}.
Some results have suggested that balanced logic styles resistant to dynamic power analysis can actually heighten vulnerability to static power SCA~\cite{Bellizia16, Bellizia17uni}.
Although balancing countermeasures have not been tested against other static SCA attacks such as LLSI and IA, we anticipate that these attack approaches are similarly capable of circumventing the protections.
Both of these SCA techniques are able to observe the leakage of bits as a function of their position in the circuit.
If optimal resolution is achieved, adversaries may simply identify which stored bits correspond to the actual secret state while ignoring their counterparts used for balancing, rendering most such approaches void. 

Time-enclosed logic~\cite{Bongiovanni15, Bellizia18, Bellizia21, Bellizia20} extends balancing approaches to encode data in the time domain such that information is only present for evaluation in a limited \emph{evaluation phase} time window during each clock cycle.
This requires an internally derived secondary clock signal to complete the evaluation phase. 
These countermeasures have the effect of shifting information leakage to higher frequencies as additional switching activity is performed.
\citet{Bellizia21} evaluated a time-enclosed logic style against static power SCA and found it to be effective in eliminating leakage from combinatorial gates, however, information still leaks from state elements.
Since these state elements are the targets of LLSI and IA, its security against these attacks is equally questionable as for the previously mentioned balancing countermeasures.

Secure implementation of masking and/or logic balancing countermeasures imposes substantial overhead.
A comparison of PRESENT\footnote{A lightweight symmetric block cipher, less resource-intensive than \texttt{AES}.} co-processor ASIC implementations with various countermeasure suites on 28\,nm technology~\cite{Moos21} finds an overhead of \texttimes2.58-3.17 (258-317\%) from first-order masking in terms of total power consumption, and \texttimes2.85-3.78 area overhead. 
Similarly for Exhaustive Logic Balancing, which is the current state-of-the-art balancing scheme to thwart static attacks, power overhead is \texttimes4.63-6.38 and area overhead is \texttimes7.97-8.38. 
Combining such schemes then becomes significantly expensive. 
As discussed before, even at such a high price it is 
unlikely that the combination of masking and balancing leads to strong resistance against all static SCA attacks, including LLSI and IA.

Randomising the order of independent operations~\cite{Veyrat12} can significantly heighten SCA difficulty as adversaries, even with knowledge of the targeted implementation, will not know exactly which sensitive values are being computed or stored at any given moment.
Several other proposed countermeasures \cite{Zhu13, Yu17, Yu_17} generate additive noise within the system to reduce the signal-to-noise ratio in power leakage.
Randomisation-based approaches are generally not expected to provide sufficient protection against static SCA, as the localisation and noise reduction features of such attacks are potentially strong enough to negate their impact (e.g., by simply reading the digital values that determine the current order of operations in a shuffling countermeasure).

In summary, with more attention having gone towards dynamic attacks, similarly countermeasures have mostly been evaluated by their resistance to dynamic leakage. 
Despite their high cost, they offer limited protection against static SCA attacks.
While some protections against static power SCA have been proposed, they can also still be overcome.
Moreover, no countermeasure (to the best of our knowledge) has yet been designed to thwart the entire family of static SCA attacks.

\subsection{Clocking in Digital Systems}
In the context of digital systems, a clock is a signal that oscillates between a high (1) and low (0) state at a given frequency.
Clocks coordinate the actions of digital systems; sequential (or `synchronous') logic elements (e.g.\ registers) in a system change state in response to active clock edges.\footnote{Typically systems use the rising edge (transition from 0 to 1) as the active edge, falling edge (1 to 0) can be used too, or a combination of both.}

\parhead{Phase-Locked Loops.}
Clock signals must be distributed to all synchronous elements in a circuit.
To drive this high fan-out clock network load in large circuits, digital designs will typically generate internal oscillatory signals for distribution and synchronise them to a reference clock generated by a crystal oscillator, which is highly accurate and stable. 
A Phase-Locked Loop (PLL) is the closed-loop control system used to achieve this synchronisation. 
PLLs are found in many digital systems within their clock management primitives, they are commonly used for clock generation, timing distribution, clock recovery, frequency synthesis, and frequency demodulation. 

\begin{figure}[h]
    \centering
    \includegraphics[width=0.95\linewidth]{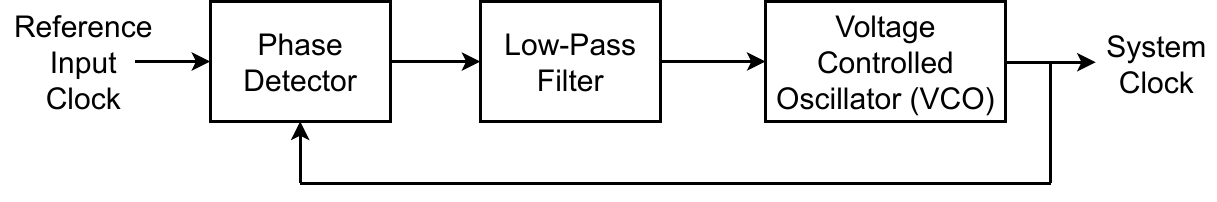}
    \caption{Block diagram showing basic elements of a PLL.}
    \label{fig:PLL}
\end{figure}

The basic structure of a PLL is shown in~\cref{fig:PLL}.
The system clock distributed throughout the circuit is generated by a Voltage Controlled Oscillator (VCO), whose frequency depends on the input voltage.
The VCO output is fed back to a Phase Detector comparator. 
This phase difference acts as an error signal fed forward to the VCO so that it adjusts its frequency to match the reference.
A Low-Pass Filter removes unwanted high frequency noise.

\parhead{Clock-Gating.}
While certain portions of a circuit are inactive, driving clock signals to them causes an unnecessary drain on power.
In a common design practice known as \emph{clock-gating}~\cite{ClockGating}, extra circuitry is incorporated in a system to dynamically enable clock distribution only to modules that are active. 
This significantly reduces the dynamic power dissipation of the gated circuit with relatively low overhead of implementing extra gating logic.
This design practice is also automated by EDA tools for optimisation~\cite{vivug, Siemens} and can thus occur without the designers knowledge. 
In ASICs not only can the clock be gated, but circuits that are not in use can be switched off~\cite{Roy03}.
This is known as power gating and it additionally reduces static power dissipation in a system. 

Clock-gating has been shown to reduce power consumption by up to 40\% in smartcards~\cite{Zhou14} and up to 50\% in mobile Systems on a Chip (SOCs)~\cite{Lee23}.
We note that PLLs cannot work inside clock-gated circuits as PLLs require a long-term stable incoming clock before they can achieve synchronisation.

\parhead{Clocking in Side-Channel and Fault Injection Attacks.}
Another popular class of attack that involves clock signal manipulation are clock glitching fault injection attacks~\cite{Alshaer2022, Alshaer2023}. 
In these attacks rapid glitched clock pulses are injected between regular pulses which violate a circuit's timing constraints. 
This can be used to bypass certain security-critical operations or instructions by not giving them sufficient time to terminate between pulses.
\citet{FaultyClock} simulate a proposed method for detecting high frequency clock signal fault injection.
They maintain a higher frequency clock in the target and use it to sample the incoming reference signal.

\parhead{Clock Sensors.}
In safety-critical applications, Clock Failure Detection (CFD) systems~\cite{Steininger19, MicrochipCFD} are often implemented for fault resilience.
Their purpose is to detect if an incoming clock (provided by an external source such as a crystal oscillator) has failed and typically this indicator is used to signal the system to switch over to an internal clock source for long enough to carry out any necessary emergency actions. 
The failure model for CFD systems is not generally geared to security and does not account for disruption by malicious actors with full knowledge of the design and complete physical system access.

In the context of security, \citet{Kommerling99} discuss incorporating a robust low-frequency (clock) sensor in the context of protecting smartcards from bus observation using a Scanning Electronic Microscope.
Lowering the target's clock frequency is said to make this kind of e-beam testing easier. 
They suggest that implementations of filter elements used to detect low-frequency input clocks are commonly found in smartcard processors but are inadequate, and they highlight that defences should be designed to be embedded within processors.
\citet{Farheen23} design a clock-freeze detection sensor alongside a voltage modulation sensor as a proposed countermeasure to 
LLSI~\cite{Krachenfels21}, they claim it may also be applicable to static power SCA attacks.
However, this detection mechanism has only been implemented in isolation and has not been evaluated within a larger design, or against any attacks.
Moreover, their design does not allow operation within a clock-gated system since their detector takes several cycles to de-assert its detection alarm flag. 

\section{Threat Model}
\label{section:threatmodel}
As in past works \cite{KocherJJ99, MangardOP07}, we consider a threat model wherein an attacker has physical access to the target device and full knowledge of its underlying implementation.
The attacker aims to leverage an idle state where the clock signal provided to the target circuit, which may only be a small component within a larger device, is stopped at a particular cycle.
During such an idle state, sensitive data remains present in state register elements (i.e.\ flip-flop cells) for the duration of an extended measurement period, long enough to extract information through static leakage.
For the non-power static SCA attacks (LLSI and IA) this period requirement is demonstrated to be in the order of milliseconds.
For static power SCA, in \cref{section:memory} we find it to be in the order of hundreds of microseconds.  
The idle state can manifest in two ways:
\begin{enumerate}[nosep,leftmargin=*]
    \item Attacker-controlled clock (\cref{fig:topdiagram} right panel): IC packages in embedded systems will commonly expose a clock pin, so in many cases this is a trivial extension of the physical access requirement already imposed. 
    This likely requires some extra work such as interrupting a circuit board connection between the target and its external clock source to connect an attacker-controlled source in place. 
    \item Naturally stopped, gated clock (\cref{fig:topdiagram} left panel): lightweight systems typically employ clock-gating for its substantial power savings. 
    Such systems usually involve a master circuit that clock-gates (deactivates the clock provided to) a target circuit. 
    For example, a master processor only activates the clock signal for a cryptographic co-processor when tasking it to encrypt some data, and at all other times the co-processor is clock-gated. 
    Concerningly, clock-gating can be incorporated in systems by design tool optimisers without the designers knowledge. 
\end{enumerate}

The target can also implement a suite of state-of-the-art SCA countermeasures, e.g.\ from the masking and/or hiding domains. 
In the case of a cryptographic target we assume that the fundamental key storage is secure, be it hardened or naturally resistant to direct key readout attacks from optical and physical probing as well as impedance and power measurements.
Otherwise, it would be simpler to directly extract keys from storage than to perform SCA.
This secure storage may be realised for example through Secure Non-Volatile Memory (SNVM) or device fingerprint extraction using a weak Physically Unclonable Function (PUF).

\section{\bt Countermeasure}
\label{section:countermeasure}

We propose an in-chip countermeasure called \emph{\bt}, to be deployed within a target circuit, that addresses the above threat model and eliminates exploitable leakage of sensitive data under any form of stopped clock condition.
\bt involves equipping a target design with circuitry that monitors the incoming clock signal for a stop condition.
Upon detection, the countermeasure triggers an alarm that clears any sensitive (e.g.\ key-dependent) information registered, 
thereby eliminating the exploitable source of static data leakage.

We propose two approaches for detecting a stopped clock, both of which are applicable to ASIC and FPGA implementations.
The first uses standard clock management circuitry that contains a PLL, as found in many conventional digital designs.
The second solution involves a custom asynchronous delay-based clock-sampling module.
This can be used in conventional designs as well as in clock-gated systems, where PLLs will not work. 
For each approach we also propose minimal alteration of the target circuit implementation such that internal registers storing sensitive intermediate data are immediately cleared 
upon a detection alarm, with the aim of preventing information leakage of those intermediates during periods of clock inactivity.

Outside of incorporating \bt we recommend any target circuit clears its sensitive registers between operations, e.g.\ after encryptions/decryptions. 
We also recommend signalling the alarm to any master circuitry (e.g.\ `data invalid' signal) to prevent false positives from affecting the overall system's functional correctness.

The means by which registers are cleared could be as simple as tying the detection alarm signal to the asynchronous reset input of sensitive registers. 
These resets set registers to zero without needing an active clock edge.
However, we note that this would invoke a number of 1-to-0 transitions equal to the hamming weight of the stored sensitive value, so clearing in this manner may increase dynamic leakage of sensitive values, even in cases where the registered values are masked. 
To address this, we propose the implementation of a \emph{masked clear}, a means of clearing where sensitive values are replaced by randomness~\cite{Bhasin2010, Moos_2020},
similarly performed by the OpenTitan\footnote{An open-source, state-of-the-art hardware root-of-trust.}~\cite{opentitanaes} cryptographic modules.
For this we assume the availability of a cryptographically secure RNG that can provide the randomness needed at any crucial clock cycle. 
The RNG can take the same clock input as the target, as it will present the fresh randomness needed at its output for the masked clear even when its clock source has stopped.

Our countermeasure is implemented at the logic design stage and requires no alternative logic style and no alteration in the standard cell libraries. 
We now describe our two approaches for implementing \bt in greater detail. 

\subsection{PLL-Based Detection}
\label{subsection:ngated}

For targets that receive a stable clock signal our clock monitoring solution is based on the function of a PLL.
The main advantage of this design is that it relies on existing, well-understood engineering solutions that provide robust clock management. 
The other advantage is that minimal design effort is required. 
PLLs are commonplace in larger designs, so in some cases building our proposed protection will only involve adding the extra components outside of the PLL.
Otherwise the extra engineering for implementing this solution is still minimal, incorporating PLLs is straightforward as established solutions are widely available for deployment on most varieties of FPGA and ASIC platforms. 

When a PLL's reference (input) clock and internally generated clock (output provided to the system) are frequency- and phase-matched, the PLL is said to be \emph{locked}.
To ensure synchronisation, sequential logic elements clocked by a PLL can be held in a reset state until lock is achieved. 
For this reason clock management modules that contain PLLs usually provide hardware designers with the option of using output status signals, such as a \texttt{LOCKED} signal, to indicate lock synchronisation.
The time a PLL needs to attain a locked state is known as lock time, and is a crucial design parameter among PLLs, generally in the order of microseconds~\cite{kintex7, IntelMAX10}. 

We propose using the logical NOT of the \texttt{LOCKED} signal ($\overline{\texttt{LOCKED}}$) as our clock monitor.
PLLs are typically sensitive enough to de-assert \texttt{LOCKED} within one clock cycle of a stopped reference signal~\cite{clkug}.
Therefore, by using such a signal to trigger a register clear we can reduce the time in which data-dependent power consumption is observable to only one immediate additional clock period.

\begin{figure}[h]
    \centering
    \includegraphics[width=0.9\linewidth]{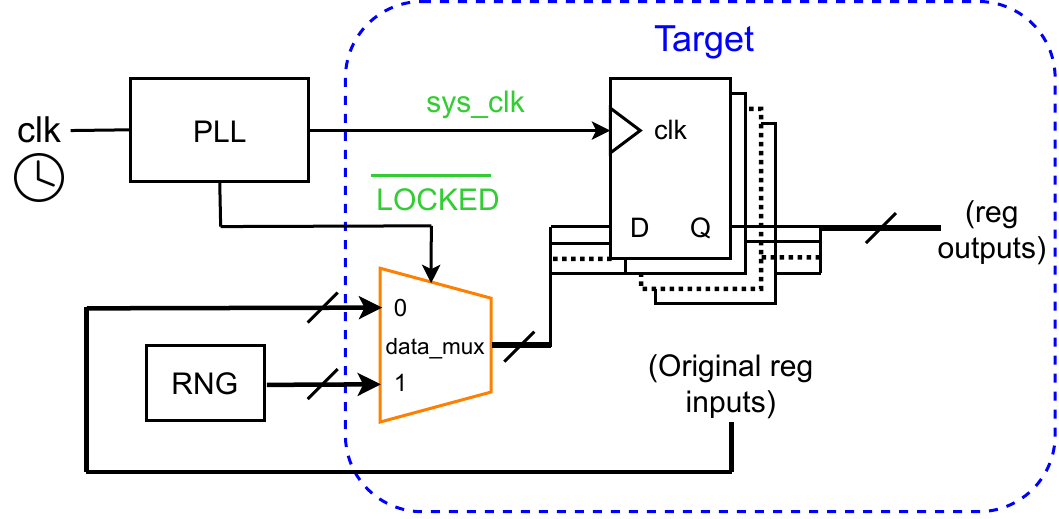}
    \caption{PLL-based \bt countermeasure system design. Upon detection of a stopped clock, the alarm signal $\overline{\mbox{\texttt{LOCKED}}}$ selects random values to be written into sensitive registers, performing the masked clear. 
    } 
    \label{fig:PLLsolng}
\end{figure}

In order to clear sensitive register contents we propose using the alarm signal ($\overline{\texttt{LOCKED}}$) to select a random number generator (RNG) as the input source of the registers, see \cref{fig:PLLsolng}.
We highlight \texttt{data\_mux} in orange because it is the only internal modification of the target circuit required, aside from additional input signals. 
The registers require an active clock edge to latch the updated inputs.
For this, in some cases PLLs will briefly continue to provide a ticking output that we can leverage.
Otherwise we can use the delayed pulse mechanism which we describe in the next design variant.
So long as the ($\overline{\texttt{LOCKED}}$) signal is not tied to the register resets (as it would be if we wanted to perform a clear-to-zero instead of masked clear) nor to the RNG reset, the first active clock edge will clear sensitive registers with random values. 
This is a similar operating principle to that of CFD systems, where an internal oscillating signal is fallen back on to execute operations allowing safe (or in this case secure) failure. 

\begin{figure*}[h]
    \centering
    \includegraphics[width=\linewidth]{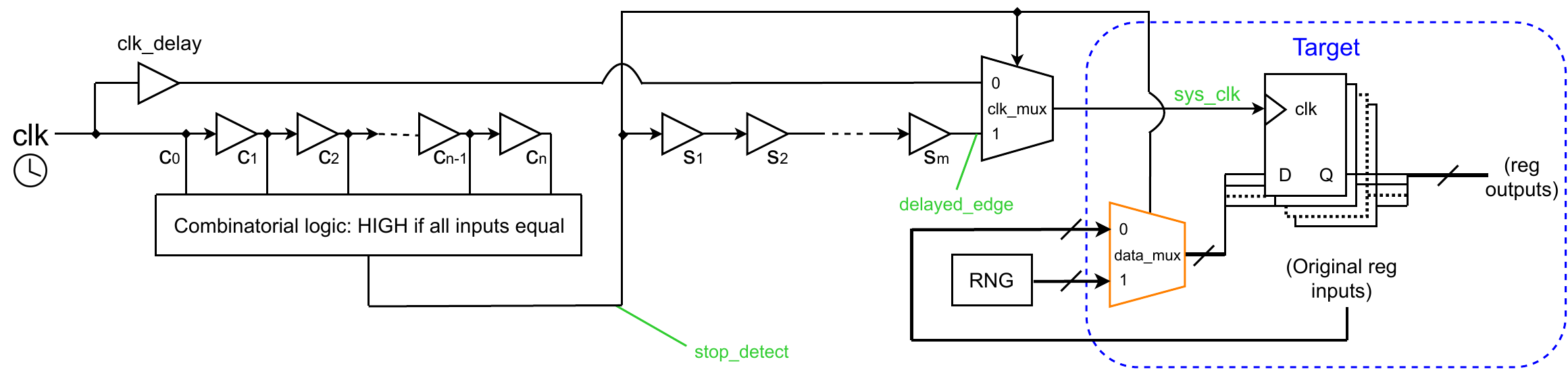}
    \caption{Asynchronous delay-based \bt countermeasure system design.
    Upon detection of a stopped clock, the alarm signal \texttt{stop\_detect} selects random values to be written into sensitive registers. 
    \texttt{delayed\_edge}, a delayed version of \texttt{stop\_detect}, is instantly selected by \texttt{stop\_detect} as the target system clock and provides an active clock edge to latch random input values, performing the masked clear.
    } 
    \label{fig:gatedsoln}
\end{figure*}

The main limitation of the PLL-based solution is that it is not applicable in clock-gated circuits because in order to operate correctly, PLLs rely on a stable incoming clock signal over a long period of time.
Introducing a PLL inside a clock-gated circuit would cause long start-up times, that are incompatible with the need to activate such circuits on-demand.
Another limitation of this solution is the amount of power drawn by PLLs which is relatively large compared to the demands of a lightweight logic core.
This makes the PLL solution far less suitable for lightweight, low-power systems.

\subsection{Asynchronous Delay-Based Detection}
\label{subsection:gated}

Given the limitations of the PLL-based solution, 
we now propose a lightweight alternative. 
The main advantage of this alternative is that it is low-power, both from the overall circuit overhead, and from the fact that it can work in clock gated-architectures. 
Thereby it allows the system in which it is deployed to benefit from substantial further energy savings.
Since we cannot rely on a continually transitioning clock, we build this alternative from asynchronous circuitry, i.e.\ digital circuitry that does not use a global clock for synchronisation.
It is applicable to ASIC and FPGA systems, however it requires significantly more engineering effort than the PLL-based solution.
Though it is ultimately implemented at the logic design stage, this solution requires a careful design process informed by multiple development stages, from logic design down to physical layout. 
Moreover, designs for certain underlying CMOS technologies are not directly portable to others.
Porting the design between technologies at minimum requires tuning it for the target platform's physical characteristics. 

The overall goal of our custom system is to overwrite sensitive register values with random data upon detection of a stopped clock signal, otherwise allowing the target to operate as normal without interrupting its operation.
At a high level, we require two main mechanisms.
We first need to detect that a clock has stopped, and then we need to generate another clock edge that will allow us to 
clear the registers.
We note that our system fails securely in the event of false positives, performing an unintended clear. 
Our system is inherently disruptive, i.e. an encryption cannot be completed if \bt is triggered during its execution. 
Therefore we prioritise minimising false positives to not adversely affect performance.
Considering the threat model, we must also design for robustness in the face of adversarial manipulation. 
We now describe how we implement these mechanisms and then discuss important design considerations and how we avoid triggering false positives.

\parhead{Clock Monitoring.}
The clock-monitor circuit consists of a chain of 
unit delay elements that takes the incoming clock signal  
as the input.
See the left side of \Cref{fig:gatedsoln}.
The output of each element is a version of the clock, time-shifted by the propagation delay time through the delay element.
A subset of these time-shifted versions of the clock are fed as input to a combinatorial logic circuit that goes high if all of its inputs are equal. 
If all inputs to the combinatorial gate are zero, the clock has been stopped low; if they are all one, it is stopped high.
In both of these cases the \texttt{stop\_detect} signal goes high, otherwise it stays low.
We refer to each input signal line fed to this circuit as a \emph{tap} on the chain of delay elements.

\begin{figure}[h]
    \centering
    \includegraphics[width=0.87\linewidth]{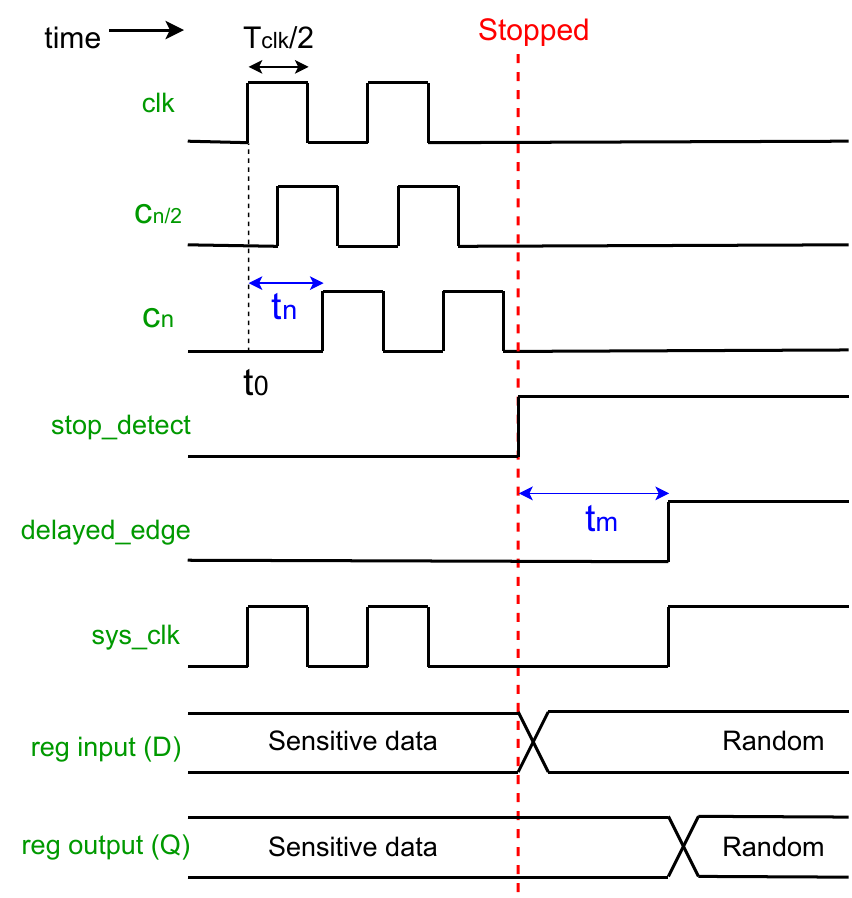}
    \caption{Timing diagram of system signals with a stopped clock condition.}
    \label{fig:gatedtiming}
\end{figure}

\parhead{Generating an Active Clock Edge.}
Upon detecting that the clock has stopped, we need to clear sensitive registers.
Again, we note that a naive clear-to-zero can be performed by tying \texttt{stop\_detect} to registers' asynchronous reset.
To instead do so with a masked clear, we follow the approach of the PLL-based solution and multiplex the registers' input (\texttt{D}) using \textit{data\_mux} such that RNG output is selected by \texttt{stop\_detect} upon a stopped clock condition.
However, unlike the usual case of the PLL-based solution we can no longer rely on a continuously generated clock signal to drive the register clear,
because the stopped input clock will not generate active edges.
Instead, we also multiplex the clock signal fed to the registers (\textit{clk\_mux}) to select a source signal which we can guarantee to provide an active edge (low to high) transition while the actual clock source is held (stopped) high or low.
For this we feed in a time-delayed version of the \texttt{stop\_detect} signal, labelled \texttt{delayed\_edge}.
\cref{fig:gatedtiming} shows the entire process in a timing diagram. 
Importantly, we note that while the figure only shows the case that the clock is stopped low, the countermeasure works the same if stopped high.
The reason being that immediately after `Stopped', \textit{clk\_mux} will assign \texttt{delayed\_edge} (which will be temporarily held low) to drive \texttt{sys\_clk}.

\parhead{Timing Design.}
Before describing the technical details of this solution we first establish some notation.
We label the $i^{\mathit{th}}$ time-shifted clock signal on the delay chain as $c_i$ and denote the propagation delay from the original clock signal $clk$ to $c_{i}$ as $t_i$.
Our solution effectively samples the incoming clock signal concurrently at $n$ different points in time from the current clock to the clock $t_n$ seconds earlier.
$t_n$ is the time-shifted version of the clock with the longest delay from the original.

Our system relies on observing variation of the clock signal as an indicator that the clock is operating normally.
Therefore, the span of time across which we sample the clock must be long enough to capture its variation, otherwise the system will trigger false positives.
This gives a minimum bound of $t_n > T_\mathit{clk}/2$, where $T_\mathit{clk}$ is the nominal clock signal period. 
\cref{fig:gatedtiming} depicts a near-minimum-allowable $t_n$. 
It shows the temporal relationship between three different versions of the clock signal: the original $clk$, the time shifted $c_n$ with the maximum delay ($t_n$) from the original, and one of the in-between delayed clock signals $c_{n/2}$ with approximately half of the maximum delay. 
At any given moment in the time window from $t_0$ to immediately before time `Stopped', the values of the three versions of the clock are not all equal, which means that 
at these times, the combinatorial circuit does not trigger an alarm, allowing the target circuit to operate as normal.

At time $t_n$ after the clock (\emph{clk}) has stopped, all versions of the clock agree, and the combinatorial circuit detects the situation (marked `Stopped' in \cref{fig:gatedtiming}).
If $t_n$ were less than $T_{\mathit{clk}/2}$, then a false positive would be triggered in our combinatorial circuit because all of the sample signals would be 0 immediately before the second clock pulse in $clk$, despite the clock actually still operating normally at that time.

$t_n$ also sets how quickly our system can detect a stopped clock.
In \Cref{fig:gatedtiming} the final transition (1 to 0) of $c_n$ occurs $t_n$ seconds after the same transition for $clk$, and only then is the stopped clock detected.
While the absolute limit is that $t_n$ must be greater than half the clock period, a larger $t_n$ should be used in implementation for robustness in the face of clock jitter and possible variations in the chain's propagation delays.

Another potential source of false positives is aliasing, 
which can occur when the sampling frequency is too low, i.e.\ below the Nyquist Limit.\footnote{$f_s > 2f_{\mathit{signal}}$, where $f_s$ is the sampling frequency and $f_{\mathit{signal}}$ is frequency of the signal being observed.}
A simple example of how this can trigger a false positive is if our sampling frequency is exactly the same as the incoming clock frequency.
In such a case, each sample point would be at the same phase of different clock cycles 
thus signal variation would not be captured.
We define the sampling frequency of our system using the shortest delay between any successive taps on the delay chain that are fed into the combinatorial logic element, i.e.\ $f_s = 1/min(t_{i} - t_{i-1})$.
To avoid aliasing $f_{clk} < f_{s}/2$.

When \texttt{stop\_detect} goes high, the input of sensitive registers switches to the RNG.
This change must be stable long enough before the active \texttt{delayed\_edge} arrives to clock the new value into the register, such that the hold time requirements of the registers are met.
Since typical register hold times are very small compared to clocking periods, setting the secondary chain $(s_1, ..., s_m)$ total delay $t_m$ to be comparable with half a clock is a safe choice.

\parhead{Clock Glitching and Gated-Restart.}
Our design is
tolerant of attacker-induced clock glitching with the inclusion of some delay (\texttt{clk\_\allowbreak delay}) on the clock line fed into \texttt{clk\_mux}. 
Without this, an attacker could send a series of well-timed glitched clock pulses which would set \texttt{stop\_detect} low while no clock transitions reach the target circuit. 
We elaborate on this scenario in \cref{app:robustness}. 
In design, \texttt{clk\_delay} must be greater than the delay through the combinatorial element (e.g.\ from \texttt{\texorpdfstring{c\textsubscript{0}}{}} to \texttt{stop\_\allowbreak detect}).
Upon circuit reactivation, using the delayed clock also enables the circuit to work from the first active clock edge following a stopped/gated period as \texttt{stop\_detect} will be de-asserted before the edge reaches the target. 
This introduces some clock skew between the target and master circuits that must be accounted for in design. 

\subsection{Comparison of Proposed Designs.}

In \cref{table:proscons} we summarise the pros and cons of each design.

\begin{table}[ht]
  \small
\centering
  \begin{tabular}{@{}lccc@{}}
 &  \textbf{Low-power} &  \textbf{Clock-gating} & \textbf{Low design cost} \\
  &   &  \textbf{compatible} &  \\
\toprule 
\textbf{PLL} & \ding{53}  & \ding{53} & \ding{51} \\
\textbf{Asynchronous} & \ding{51} & \ding{51} & \ding{53}\\
\bottomrule
\end{tabular}
\vspace{0.5em}
  \caption{Pros and cons of \bt variants.}
\label{table:proscons} 
\end{table}

\subsection{Potential Design Automation}
Most aspects of applying \bt to different targets could lend themselves to automation. 
Implementing a PLL for the detection mechanism already requires low engineering effort as it only involves instantiating an IP, which in many cases may be further assisted by \emph{Clocking Wizard} tools typically provided by FPGA design suites. 
Tuning the delay chain for the alternative detection mechanism can similarly be automated given the availability of characterisation information about the delay of circuit elements. 
All that designers would need to decide on is the operating frequency and tolerable false positive margins.
Otherwise, the Trivium RNGs are flexible to produce as many random bits as needed per cycle. 
The only remaining problem is the identification of registers that store sensitive values at any point during execution.
For expansive designs a manual process could be cumbersome and prone to mistakes.
For such situations, we recommend using automated tools such as REBECCA~\cite{Bloem18, rebecca} to assist the process.
Using REBECCA, designers can annotate circuit inputs with labels `sensitive', `mask', or `public'.
The tool propagates these throughout the circuit, ultimately informing the designer which registers store sensitive values at any point during execution.

\section{Evaluation Setup}
\label{section:attack}

To evaluate our \bt countermeasure, we perform practical static power SCA attacks using various attack setups and against different target cryptographic hardware devices, attempting to extract their secret keys with and without \bt in place.
We perform Correlation Power Analysis (CPA)~\cite{Brier04} attacks against \texttt{AES} and masked \texttt{SKINNY} implementations.
We describe CPA against \texttt{AES} and \texttt{SKINNY} in detail in \cref{app:cpaaes,app:cpaskinny}, respectively.
Here we describe the targeted implementations, our specific measurement setups for attacking each, then the general procedure for mounting attacks.
Lastly, we carry out end-to-end attacks against the unprotected targets to validate the setups.

\subsection{\texttt{AES} Target without SCA Protections on 7 Series FPGA}

\label{subsec:target}
Our first target is a third-party \texttt{AES} core~\cite{saseboAES} configured onto a Xilinx 28\,nm \mbox{Kintex-7} FPGA.
The FPGA is housed on a \mbox{SAKURA-X}\footnote{Also known as the \mbox{SASEBO-GIII.}}~\cite{Hori12} board, which features dedicated measurement points across a shunt on the FPGA's power supply line. 
The implementation has no SCA protections and uses a round-based architecture (one round is performed per clock cycle). 
We purposely attack an implementation without standard side-channel protections and under optimal conditions to bias in favour of the attacker and best illustrate the contrast in ease of key extraction without vs.\ with \bt deployed. 

\parhead{Point-of-Interest.}
We mount a final-round attack as described in \cref{app:cpaaes}, 
where the intermediate key-dependent values 
we aim to recover are the inputs to the final round SBoxes. 
In our target these are registered between the penultimate and final active clock edges. 
We stop the clock during this cycle. 

\parhead{Measurement Equipment.}
We derive the current drawn by our target from voltage measurements across its shunt resistor.
We measure this with a LeCroy AP 034 Active Differential Probe attached to a LeCroy Waverunner 6100A oscilloscope. 

\parhead{Temperature Control.} 
For optimal attack conditions we place the target inside a climate chamber, the utility of which is two-fold.
Static leakages are amplified at higher temperatures~\cite{Moos20}, and the controlled environment stabilises some of the influence that temperature variations have on leakage. 
We set the climate chamber to 60$^\circ$C, because the \mbox{SAKURA-X} board was found to sporadically power off at higher temperatures.

\subsection{\texttt{SKINNY} Target with first-order Masking on 6 Series FPGA}
Our other target is a third-party \texttt{SKINNY} core~\cite{AEAD} with first-order masking protection in place, using two shares.
Specifically, the implementation realises \texttt{SKINNY-128-128} encryption 
based on a SCA-protected core developed to study the leakage resistance of the NIST lightweight cryptography competition finalist Romulus~\cite{VerhammeCS22, ShenPSCV22}.
We configure the implementation onto a Xilinx 45\,nm Spartan-6 FPGA which is housed on a SAKURA-G~\cite{Guntur14} board.
We target this implementation to demonstrate that \bt can seamlessly be integrated with standard side-channel protections which, on their own, are insufficient to protect against static SCA attacks. 

\parhead{Point-of-Interest.}
We mount a second-round second-order attack as described in \cref{app:cpaskinny}, this is the first round that involves key material in computation. 
We stop the clock after the second SBox of the second round has been computed.
At this point the result is stored in registers in a masked representation while part of the multi-cycle pipelined Sbox circuit (see~\cite{VerhammeCS22}) is also still holding related information.
This maximises the amount of leakage that can be exploited.

\parhead{Implementation Details.}
The masked \texttt{SKINNY} core is built based on trivially composable gadgets, following the Hardware Private Circuits (HPC) masking scheme~\cite{Cassiers21}.
The authors have formally verified its security using the fullVerif tool~\cite{Cassiers21}.
Additionally, we have practically verified the absence of first-order leakage using 100 million dynamic power traces and 1 million static power traces using a fixed-vs-random t-test from the standard Test Vector Leakage Assessment\footnote{A method of evaluating device leakage without performing an attack.} (TVLA) methodology~\cite{SchneiderM15}.
Therefore, all the following attacks we present on this target are based on second-order analysis.

\parhead{Measurement Equipment.}
In this setup we directly measure the current drawn by our target.
We connect it to a Keithley 2450 Source Measure Unit (SMU) which serves a dual purpose of supplying power along with capturing high-accuracy measurements of the current drawn.
The same kind of device has been used in previous literature~\cite{Moos21,PrimeMasking}.

\subsection{Measurement Procedure}
\label{section:inj}

\parhead{Input Control and Measurement Interval.}
When we stop the target clock at the point-of-interest for our given measurement interval we also set all other inputs to zero.
We must stop the clock long enough for the memory effect to subside before we start taking measurements, and we need a sufficiently long measurement period for intra-trace averaging to remove a significant amount of noise.
We refer to the length of time from when we stop the clock to the beginning of measurement acquisition as the \emph{offset}, and to the time from start to end of the acquisition period as the \emph{window length}.
Past works~\cite{Moos17, Moos20} have indicated 20\,ms as a good rule of thumb. 
In our attacks we use a 25\,ms offset and a 20\,ms window.

\parhead{Pre-Measurement Warm Up.}
We find that internal (in-chip) temperature effects have considerable impact on leakage measurements. 
Performing computations generates heat within the targets.
To stabilise this effect we warm up the chip internally by performing the same computations (i.e.\ encryptions) at the same rate as then performed in the measurement phase. 

\parhead{Post-Processing Temperature Control.}
When placed either in a room or 
placed inside a climate chamber, the targets are subjected to temperature drifts. 
These effects are exhibited in static power traces mostly as slow drifts up and down, whose variation far exceeds that between successive traces. 
Previous works mitigated this noise with various post-processing strategies such as moving-average high-pass filtering~\cite{Moos20} or interleaving measurements with a known baseline state and taking the differences~\cite{Moradi14}.
We have opted to remove all low-frequency signal components by applying a high-pass filter to the traces taken over the measurement period.

\subsection{Validation with End-to-End Attacks}
Our CPA attacks successfully recover keys from both targets. 
Thus, we validate our attack setups and procedures as a viable means of exploiting static power side-channel leakage.

\begin{figure}[h]
    \centering
    \includegraphics[width=\linewidth]{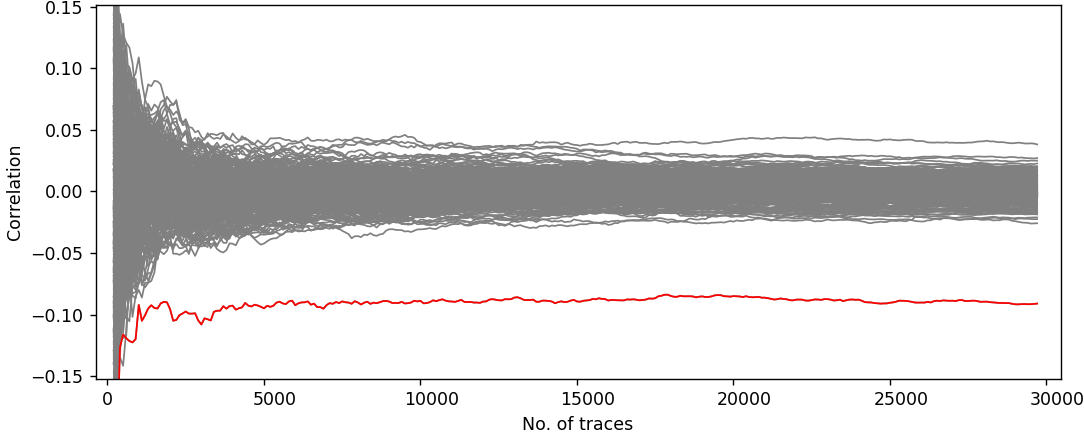}
    \caption{CPA on unprotected \texttt{AES} targeting the 10th key byte, 600 MTD.} 
    \label{fig:MTD_byte10}
\end{figure}

\parhead{Attacking Unprotected \texttt{AES}.}
\Cref{fig:MTD_byte10} shows the correlation of the static power measurements with expected power consumptions for each of the 256 potential \texttt{AES} subkeys, for the 10th byte (subkey) of the key. 
The correlation of each of the 255 incorrect candidate subkey guesses over the 30,000 traces are shown as grey lines, and the correlation of the correct subkey guess is shown by the red line. 
Here, the correct subkey candidate for the 10th key byte clearly emerges with the greatest correlation after the first 800 traces.
Note, this is from one subkey.
On average, recovering all 16 correct subkey candidates took approximately 1,500 measurement traces, so our Measurements to Disclosure (MTD)~\cite{TiriK2005PIwW} value (of the whole key) is 1,500.
We calculated this average across all subkeys by using 10 disjoint subsets of our measurement trace. 
This attack success metric offers a relative benchmark on the level of leakage under optimal conditions given our setup.

\begin{figure}[h]
    \centering
    \includegraphics[width=\linewidth]{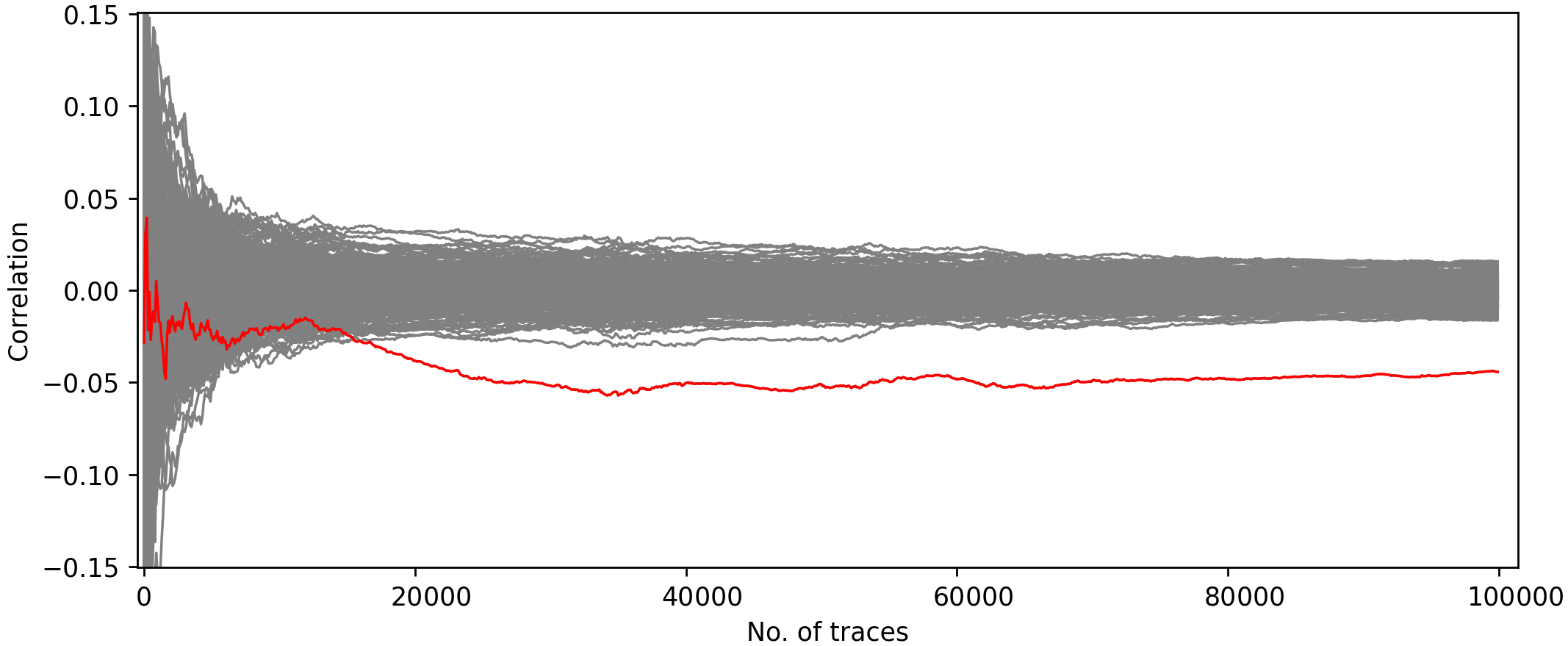} 
    \caption{Second-order CPA on unprotected (i.e.\ not equipped with \bt) first-order masked \texttt{SKINNY} targeting the 2nd key byte, 16,000 MTD.}
    \label{fig:MTD_skinny}
\end{figure}

\parhead{Attacking Masked \texttt{SKINNY}.}
Similarly, \Cref{fig:MTD_skinny} shows the results of a second-order CPA performed on 100,000 measurement traces of the first-order masked \texttt{SKINNY} implementation, targeting the second key byte of the second round.
The attack succeeds in isolating the correct key candidate after approximately 16,000 MTD.
While the first-order protection increases the amount of measurements needed to perform a successful attack (compared to the unprotected \texttt{AES}), it is evidently insufficient to resist static SCA, due to the low noise property of such attacks.

\section{Assessing the Memory Effect.}\label{section:memory}

We have described our CPA attack setups and procedures and validated them by performing key recovery attacks against both targets, also producing MTD metrics indicative of system leakage.
Now, we use the same techniques to assess the influence of the memory effect on overall attack performance 
by using acquisition windows with shorter offsets, i.e.\ the window starts closer to the final clock edge before it is stopped.
Given that the \texttt{AES} target exhibits much higher overall leakage (lower MTD), we use this setup for our memory effect analysis to continue biasing in favour of the attacker.

We recall that due to the memory effect, dynamic effects from circuit state transitions do not subside immediately once the logical transitions themselves have terminated~\cite{Moradi13}.
These effects linger and can influence power measurements taken long after the transitions~\cite{Moradi13, Moos17, Moos20}.
This is significant for \bt's effectiveness against static power SCA because it drastically increases the time window that data needs to remain stable and available for attacks to succeed.

We evaluate the memory effect by repeating the attack against the unprotected \texttt{AES} target with varying time offsets and window lengths in order to gauge the duration and extent of its influence on leakage measurements. 
We begin by fixing the offset at 20\,ms and narrowing the acquisition window over repetitions of the attack until there is a noticeable decrease in attack performance. 
We find a 250\,\textmu s window length to be the smallest with comparable attack performance to the 20\,ms window.
Performing the attack with this window length will be very sensitive to the leakage signal strength. 

\begin{figure}[h]
    \centering
    \includegraphics[width=0.95\linewidth]{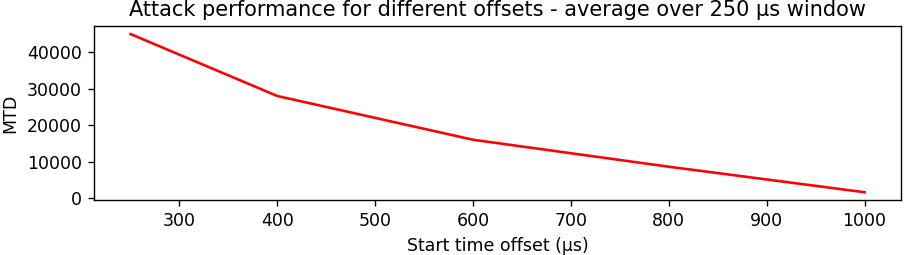}
    \caption{CPA attack performance against unprotected \texttt{AES} for various start-time offsets with 250\textmu s measurement window.}
    \label{fig:memeffect}
\end{figure}

Therefore, we then fix the window length to 250\,\textmu s and repeat the attack, shifting the time offset closer to zero each time, i.e.\ bringing the whole span of the acquisition window nearer to the last active clock edge before the clock is stopped. 
\Cref{fig:memeffect} shows the results of the attack for varying offsets.

\begin{figure}[h]
    \centering
    \includegraphics[width=\linewidth]{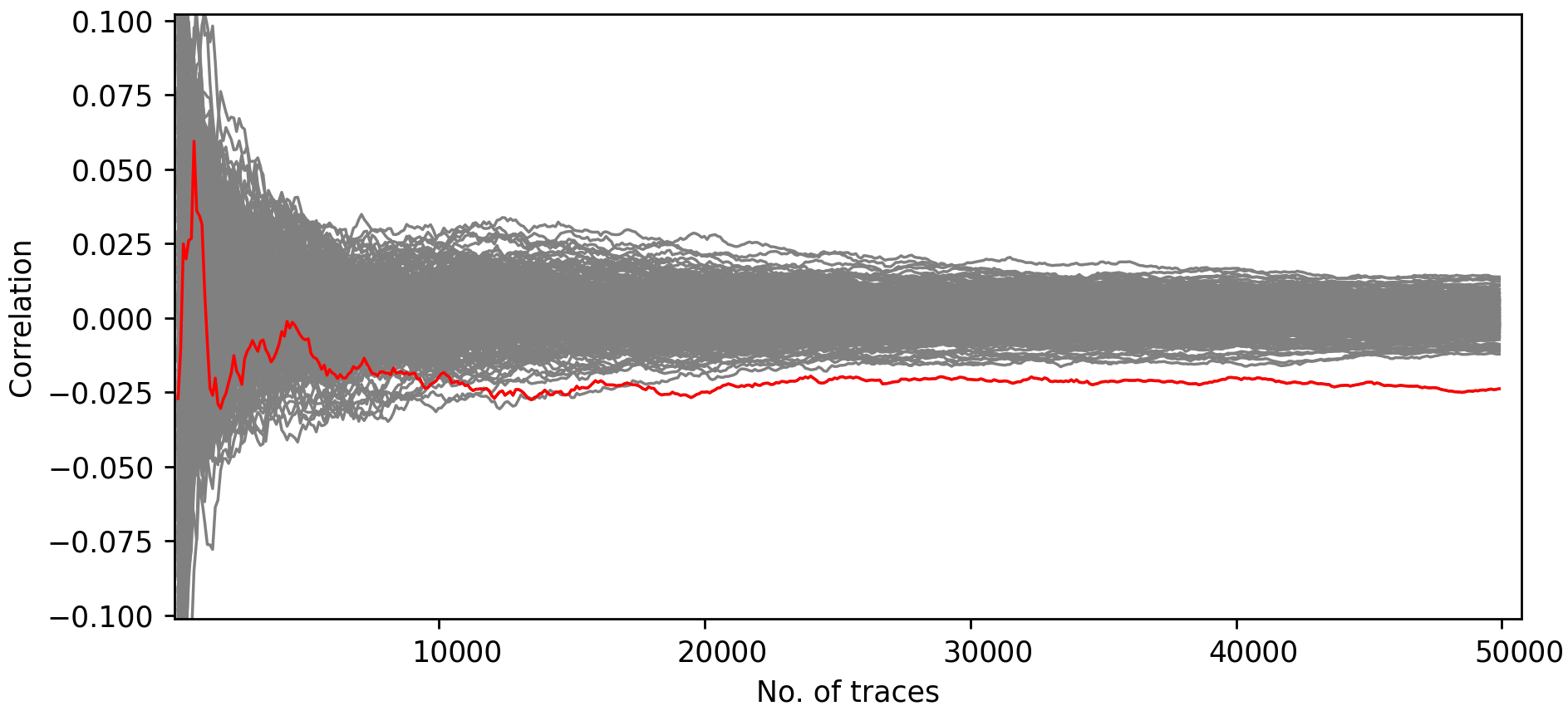}
    \caption{CPA against 10th key byte of unprotected \texttt{AES} with 400\textmu s offset, 250\textmu s measurement window, 18,000 MTD.}
    \label{fig:byte10_400}
\end{figure}

We find that the memory effect appears to influence measurements for offsets as late as 800\,\textmu s.
Offsets from 1\,ms upward all had similar performance of around 1,500 MTD.
\Cref{fig:byte10_400} shows an example CPA key guess progression with just under 20,000 MTD at the 400\,\textmu s offset.
From that point, with reducing offset the memory effect influence increases until 200\,\textmu s where the attack is unsuccessful with 100,000 measurements. 
We use this as our cut-off since it is approximately two orders of magnitude greater than the optimal MTD.
From these results, we deem the memory effect to be dominant for at minimum the first 200\,\textmu s after the last active clock edge. 

Regarding our countermeasures, if we take this to be the memory effect duration specific to our measurement setup, any configuration that detects a stopped clock within 200\,\textmu s should protect the circuit from static power SCA attacks.
We believe that the attack is increasingly difficult for shorter offsets until a point where the memory effect noise obscures leakage to the point where the attack is infeasible.
\bt (in both of its variants) can detect a stopped clock within one clock period.
This means that if \bt is deployed within a cryptographic co-processor, whose typical operating frequency would be in the MHz range, detection and clear under a stopped clock condition occurs at latest after 1\,\textmu s. 

\section{Countermeasure Implementations and Evaluation with Attacks}\label{section:evaluation}

Having described the general design and approach of incorporating the \bt countermeasure in hardware systems earlier in \Cref{section:countermeasure}, we now describe our implementations of it within the target FPGA devices.\footnote{The source RTL for both of these design artefacts is available at: \mbox{\url{https://github.com/0xADE1A1DE/Borrowed-Time}} } 
Then, using the same evaluation setups described in \Cref{section:attack} we attempt to attack the now-protected targets.

\setlength{\tabcolsep}{4.5pt}

\begin{table*}[ht]
  \small
\centering
  \begin{tabular}{@{}llcccccc@{}}
   & & & \textbf{Power Overhead (\%)} & & & & \\
 & &  \textbf{Power [mW]} &  [Total (Fixed / Variable)]  & \textbf{Critical Path [ns]} &
 \textbf{Max. Frequency [MHz]} & \textbf{LUTs} & \textbf{Regs} \\

\toprule 
\textbf{\texttt{AES}} & Base Unprotected & 116  & -- & 5.008 &
199.7 & 1387 & 535 \\
 & PLL BT Protected & 203  & 75.0 (62.9 / 12.1) & 6.009 & 
 166.4 & 2274 & 1149 \\ 
 & Async. BT Protected & 138  & 19.0 (6.9 / 12.1) & 6.009 & 
 166.4 & 4079 & 1163 \\ 
 \midrule
\textbf{Masked} & Base Unprotected & 115  & -- & 7.936 & 
126.0 & 2715 & 2722 \\
\textbf{\texttt{SKINNY}} 
& PLL BT Protected & 227  & 97.4 (80.9 / 16.5) & 7.950 & 
125.8 & 3598 & 3793 \\
 & Async. BT Protected & 135  & 17.4 (0.9 / 16.5) & 7.950 & 
 125.8 & 3697 & 3793 \\
\bottomrule
\end{tabular}
  \caption{\bt Power Overhead, Critical Path (Maximum Operating Frequency) and Resource Utilisation.}
\label{table:overhead} 
\end{table*}

\subsection{\bt Implementations}

\parhead{PLL-Based Solutions.}
To build this version of our countermeasure into our 7 Series (\texttt{AES}) target we use the Mixed-Mode Clock Manager (MMCM)~\cite{clkug} primitive.
We use an optional status output it provides called \texttt{CLKINSTOPPED}. 
For the 6 Series (\texttt{SKINNY}) target we use a PLL\_BASE~\cite{clkug6} primitive and its \texttt{LOCKED} output signal (logical NOT as the alarm). 
Both indicate a stopped input clock within one missed cycle.

\parhead{Asynchronous Delay-Based Solutions.}
To build the delay chains for clock monitoring, we are constrained to using the resources built into the target FPGAS.
As an aside, an ASIC implementation where circuit elements are customisable as with more aspects of their placement and routing would provide designers greater control over timing characteristics. 

We build the delay chains using Lookup Table (LUT) circuit primitives.
These are configurable truth tables that can implement any Boolean function to implement combinatorial logic in FPGAs.
We use the smallest such elements available for each target: 1-input 1-output \emph{LUT1}s in our 7 Series (\texttt{AES}) target, and 5-input 1-output \emph{LUT5}s in the 6 Series (\texttt{SKINNY}) target. 
Post Place and Route Simulation on delay chains built in each of our targets indicate the propagation delay between successive elements to be 190\,ps and 620\,ps on average, respectively.  
We consider these estimates to be reliable as they are the ground-truth used for timing analysis of digital designs.

To feed our clock sampling logic, in the 7 Series target we tap every $66^{th}$ element along the delay chain until around the $1900^{th}$ tap. 
We run this core at 8MHz.
This gives an average 12.5\,ns gap between taps and a total delay at the final element of 360\,ns. 
The operating frequency range this allows for is therefore 2.5\,MHz $< f <$ 40\,MHz.
In the 6 Series target we tap every $3^{rd}$ element along the delay chain until the $90^{th}$ tap.

For the remaining elements, we construct the combinatorial clock-sampling logic element using \emph{LUT6}s\footnote{The largest available Xilinx LUT primitive type by input size.} for both targets. 
For clock multiplexing we use a \texttt{BUFGCTRL}~\cite{clkug} primitive in the 7 Series and a \texttt{BUFGMUX}~\cite{6series} in the 6 Series.
These can instantly switch between two asynchronous clock sources.

\parhead{Randomness Generation.}
Following the recommendations of \citet{Cassiers23}, in all instances of \bt we implement an unrolled instance of the cryptographically secure Trivium~\cite{Trivium} with a fixed initial seed as our source of generating fresh random bits each clock cycle.
For the \texttt{AES} target, we configure it to generate 128 random bits per cycle so that the entire state register can be cleared at any point in execution.
The \texttt{SKINNY} target requires less fresh randomness at only 64 bits since there is no key-whitening in its first round, meaning the first round state only depends on plaintext. 
This allows us to accumulate the unused randomness generated during the first round in our design. 
If more random bits are needed for masked clears, 
 either the unrolling degree can be enlarged, or multiple Trivium instances can be used. 

\parhead{Runtime Overhead.}
\bt imposes no overhead in terms of additional clock cycles on what is already a round-based \texttt{AES} implementation, nor for the masked \texttt{SKINNY} implementation.
However, the insertion of the \textit{data\_mux} element, shown in orange in \cref{fig:PLLsolng} and \cref{fig:gatedsoln}, may affect a protected circuit's maximum possible clock frequency if the combinatorial path going into sensitive registers is the system's critical path. 
To evaluate the impact on our targets' maximum possible clock frequency we assess the critical path of the original (base) and the protected circuits, listed in \cref{table:overhead}.

\parhead{Power Overhead.}
Both variants of \bt introduce overhead to the protected circuit. 
We use the Xilinx toolchain to estimate the power consumption of the unprotected targets as standalone base systems, and when each variant of \bt is incorporated.
These estimates are shown in \cref{table:overhead}. 
The presented overheads incorporate all components built in for \bt, including the randomness generation circuitry. 
The asynchronous delay-based solutions are indeed better suited for lightweight, low-power systems, incurring only  19.0\% and 17.4\% overhead compared to 75.0\% and 97.4\% from the corresponding PLL variants.
Importantly, these are all drastically cheaper than state-of-the-art countermeasures which instead impose several multiples of overhead~\cite{Moos21} such as Exhaustive Logic Balancing (463-638\%), (first-order) masking (258-317\%), and their combination (1357\%). 

We break down the power overheads into fixed and variable components.
The fixed cost is from the one-off instance of the clock monitoring circuitry. 
The variable cost component scales with the number of sensitive registers (bits) protected and this is dominated by the randomness generation circuitry.
This is another aspect where \bt compares favourably to state-of-the-art schemes, as the majority portion of their overhead is variable whereas most of \bt's is fixed.
We also note that since the fixed overhead of PLL-based \bt is from the PLL itself, it would be negligible in systems that already incorporate a PLL for other purposes.

\parhead{Area Overhead.}
We list FPGA LUT and register utilisation in \cref{table:overhead}. 
We note that these numbers do not distinguish between LUTs of various input sizes, and that the LUT primitives themselves are likely to be implemented differently between targets as they are differing generations of FPGA. 

We also note that these counts are typically a measure of logical complexity whereas our constructions of \bt leverage analogue effects. 
For this reason there are several shortcomings in using these metrics to evaluate our solutions. 
We cannot make a meaningful comparison to the clock management primitives used in the PLL solutions.
The delay-chains use a large number of LUTs, particularly for the AES target.
We can significantly reduce this by replacing segments of the chain (currently composed of many LUTs) between each tap with purposefully long manually routed signal wires with equivalent propagation delays. 
While this would drastically reduce the LUT count essentially for free, the additional power overhead from driving much longer signal lines would be captured in power overhead.
Thus we consider power to be the most representative metric.

\parhead{Combination with Masking.}
In practice, designers of physically secure cryptographic systems need to protect their circuits against both dynamic \emph{and} static SCA attacks.
\bt alone cannot plausibly claim any (positive or negative) impact on the vulnerability of cryptographic implementations through dynamic (power) SCA.
However, one of its core advantages over state-of-the-art schemes is that it can be trivially combined with any common masking countermeasure to resist dynamic attacks without 
its overhead scaling with the 
order of masking used (drastically reducing relative overhead numbers compared to the ones listed above).
This is due to the fact that clearing a single share of any encoded data, by definition of the secret-sharing principle, removes all available information about the corresponding variable from the circuit.
We design our implementation for protecting the masked \texttt{SKINNY} target to clear a single share.
Applying \bt to a masked implementation instead of an unmasked one can eventually become even less costly (in relative \emph{and} absolute overhead), since 1) RNGs will already be present and 
2) a masked clear may not always be required as, depending on implementation and masking scheme, in some cases temporarily increasing the leakage of one share may not harm security.
This would permit cheaper \bt implementation with a clear-to-zero. 

Previous literature has combined masking with hiding countermeasures from the equalisation or randomisation domains to achieve strong protection against static \emph{and} dynamic attacks simultaneously. 
However, as discussed earlier, we do not expect such hiding countermeasures to be secure against LLSI and IA static attacks.
Furthermore, the masking + hiding combination (especially where hiding = balancing) has been shown to be significantly more costly than \bt applied to a masked implementation~\cite{Moos21}.
Moreover, if first-order Boolean masking (as implemented in the masked \texttt{SKINNY} target) is deemed insufficient to thwart dynamic attacks without additional hiding, higher-order masking or prime-field encodings with a proven resistance to low noise levels can be considered~\cite{Masure23, PrimeMasking}. 
The key takeaway is that we expect the combination of \bt with any form of secure masking to be significantly cheaper than other state-of-the-art combined approaches, while providing much stronger overall resistance to a wider range of advanced SCA attacks.

\subsection{CPA Attacks Against Targets Protected with \bt}

\begin{figure}[h]
    \centering
    \includegraphics[width=\linewidth]{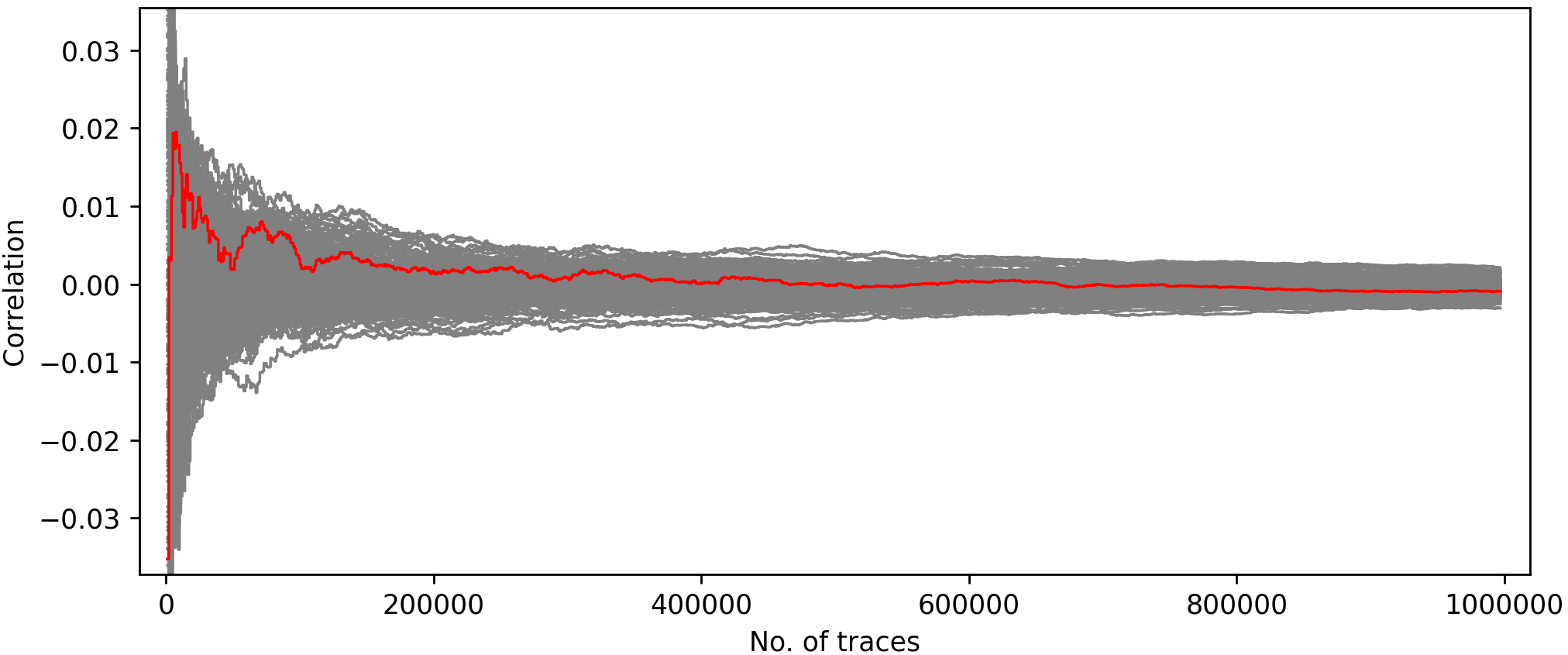}
    \caption{CPA over 1 million traces on 10th key byte of asynchronous \bt-protected \texttt{AES}, no emergence of correct candidate.}
    \label{fig:byte10_delay}
\end{figure}

To evaluate the security of \bt we repeat the attacks of \cref{section:attack}
with each variant of \bt. 
We find there to be no useful information leakage and are therefore unable to recover any subkeys with 1 million traces. 
See CPA plots \Cref{fig:byte10_delay}, \cref{app:pll} - \Cref{fig:byte10_PLL} and \cref{fig:MTD_skinny_BT}.
This is shown in the figures as the correct key guesses do not emerge with higher correlation than other candidates, instead staying around zero.
With this we confirm that the masked clears happen correctly and there is no usable residual leakage of previous data contents that can be exploited in the static power attack model.
We therefore conjecture that \bt equally protects against LLSI and IA attacks, as the sensitive information is simply not available long enough to be effectively targeted by such static SCA techniques.

\section{Limitations and Future Work}\label{section:limitations}

When \bt clears sensitive state components the computation that is interrupted can no longer be completed, making it a potentially disruptive countermeasure.
In normal operation where \bt is not triggered this will not affect runtime, however it may do so if false positive triggers occur, forcing parts of computations to be repeated. 
For example with the asynchronous \bt variant, a false positive during the execution of our round-based \texttt{AES} target would incur runtime overhead equal to the number of rounds (clock cycles) which had already completed (and would require repeating) for that given encryption. 
To account for this, we design our implementations with heavy prioritisation on robustness to false positives. 
Namely, 
we establish a wide frequency range to mitigate false positive risks of aliasing (in case of clock frequency increases) and from failing to capture clock variation (clock frequency drops). 

Although we have not produced explicit experimental evidence for it (due to the costly setup and engineering effort required), we firmly expect \bt to provide effective protection against the non-power static SCA attacks of LLSI~\cite{Krachenfels21} and IA~\cite{Monfared23} as they have hitherto been demonstrated, since both of these attacks rely on clock control for the continued presence of sensitive stored values.
However, performing these attacks may also be viable during normal execution without clock stopping, within clock-cycle time-spans.
Unlike the static power SCA scenario, such attacks might not be directly affected by the memory effect reducing the usability of measurements taken during such a time-span.
With such capability, these attacks would effectively be possible without needing a stopped clock. 
However, we expect that other factors related to required scan and sweep times to severely limit non-power static SCA attacks in such a setting. 
We leave a more careful investigation of this to future work.
These attacks are also claimed to work during extended periods when a target circuit continues to receive a ticking clock signal but does not overwrite sensitive registered values, as would be the case with a cryptographic implementation that does not clear or overwrite registers immediately after the current operation is finished.
To prevent such scenarios, we reiterate our recommendation to ensure any cryptographic implementation protected by \bt performs masked clears of its registers between encryptions/decryptions.

The scenarios considered in this work do not account for power-gated ASIC designs. 
Many power gating schemes are based on multiple-threshold circuit design~\cite{Roy03}, and of those some also allow for data retention in low-power (gated) states. 
With or without data retention, it is unclear whether such systems expose static leakage and to our knowledge there have been no demonstrations of their exploitation.
We leave exploration of this to future work. 
As with our proposed system, we recommend that power-gated systems perform masked clears of sensitive data before entering the gated state. 

Our countermeasure instances are likely to be less effective against adversaries that have highly advanced invasive capability such as the capability to edit circuits, e.g.\ by using a focused ion beam~\cite{helfmeier2013}.
With such capability, an adversary can disable \bt, e.g.\ by disrupting the alarm signal.
However, taking extra measures to route the \bt logic on the lowest metal layers can be an ASIC-based defense against such attacks.
At the same time, circuit editing facilitates other much stronger attacks, including directly observing register contents, which bypasses any need to carry out side-channel analysis.
Moreover, circuit editing equipment is quite expensive and requires a high level of expertise, significantly limiting the number of potential adversaries.

Less powerful adversaries may try to disrupt \bt, e.g.\ by using a laser to disable the alarm signal~\cite{Tajik15}. 
However, such fault injection attacks have a low spatial resolution as laser spot diameters are on the order of at minimum 1\,\textmu m~\cite{Woudenberg11}, whereas modern silicon features are typically below 100\,nm.
Hence, if system layout tightly couples the alarm signal to the target circuit, or employs some form of protective routing~\cite{Trippel23}, isolated disruption of the alarm is likely to be impractical. 
We leave validation of this to future work.

\section{Conclusion}
Side-channel attacks that exploit the vulnerability of secret data stored in register elements on integrated circuits that are temporarily in an idle mode are becoming increasingly common.
Different variants of such attacks exploiting various physical phenomena exist, including Static Power Side-Channel Analysis, Laser Logic State Imaging and Impedance Analysis.
The emergence of these attacks motivates a clear need for dedicated countermeasures.
In this work we present \bt, a simple yet effective countermeasure against static SCA attacks.
\bt operates by ensuring that registers do not contain sensitive secrets when the circuit is stopped.
We evaluate \bt using practical experiments and demonstrate that it provides an effective protection against physical attacks while imposing modest overhead.

\section*{Acknowledgments}
We thank Amir Moradi for helpful discussions and advice.
We also thank Braden Phillips, Jack Nelson, and Danny Di Giacomo for equipment and technical assistance.
Work partially done while first author, Robert Dumitru, was affiliated with the Defence Science and Technology Group, Australia.

This work was supported by
  an ARC Discovery Early Career Researcher Award DE200101577; 
  an ARC Discovery Project number DP210102670; 
  the Defence Science and Technology Group, Australia under Agreement ID10620; 
  and 
    the Deutsche Forschungsgemeinschaft (DFG, German Research Foundation) under Germany's Excellence Strategy - EXC 2092 CASA - 390781972.

\bibliographystyle{IEEEtranSN}
{\footnotesize
\bibliography{biblio}}

\appendices
\crefalias{section}{appendix}
\section{Clock-Glitching Robustness}\label{app:robustness}
As discussed in \cref{subsection:gated}, implementing asynchronous \bt without delaying the clock supplied to the target leaves it susceptible to circumvention by clock-glitching. 

\begin{figure}[h]
    \centering
    \includegraphics[width=0.87\linewidth]{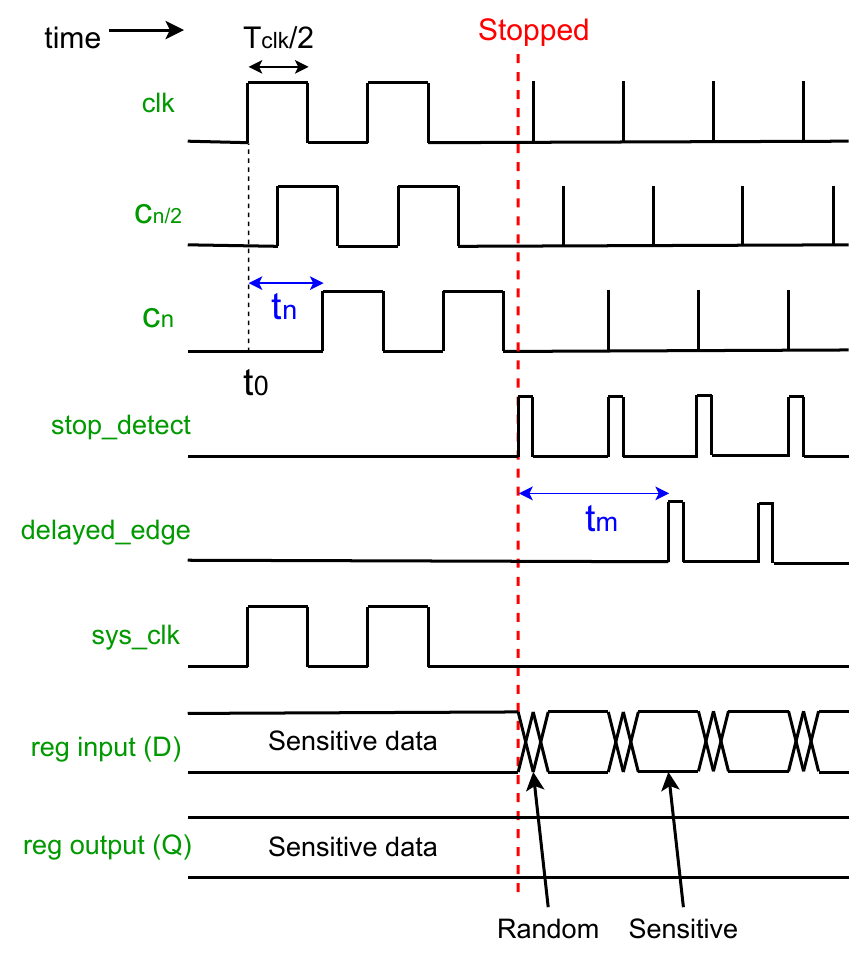}
    \caption{Clock glitching against naive implementation without delay on system clock.}
    \label{fig:clockglitch}
\end{figure}

\cref{fig:clockglitch} shows such a scenario where an attacker sends well-timed glitched pulses.
After stopping the clock, for a brief period the attacker allows \texttt{stop\_detect} to be asserted, during which time \texttt{delayed\_edge} will be selected as the system clock.
Ideally when the pulse arrives at \texttt{clk} it would go through to the system clock and provide an active edge for the registers.
However, if the delay from \texttt{\texorpdfstring{c\textsubscript{0}}{}} through to (de-assert) \texttt{stop\_detect} is longer than the pulse’s width, then in the short time that \texttt{clk} is high it will not be selected by \texttt{clk\_mux}  (so it will not provide an active edge).
This results in no edge clocking out sensitive data from target registers. 

However, using a delayed clock (as shown in \cref{fig:gatedsoln}) solves this (for example if we consider \texttt{\texorpdfstring{c\textsubscript{n/2}}{}}, although this would not be used in an implementation since it provides much more delay than needed) as the pulse will arrive when \texttt{stop\_detect} is de-asserted and therefore go through to provide the target an active edge.

\section{PLL \bt}\label{app:pll}
\begin{figure}[H]
    \centering
    \includegraphics[width=\linewidth]{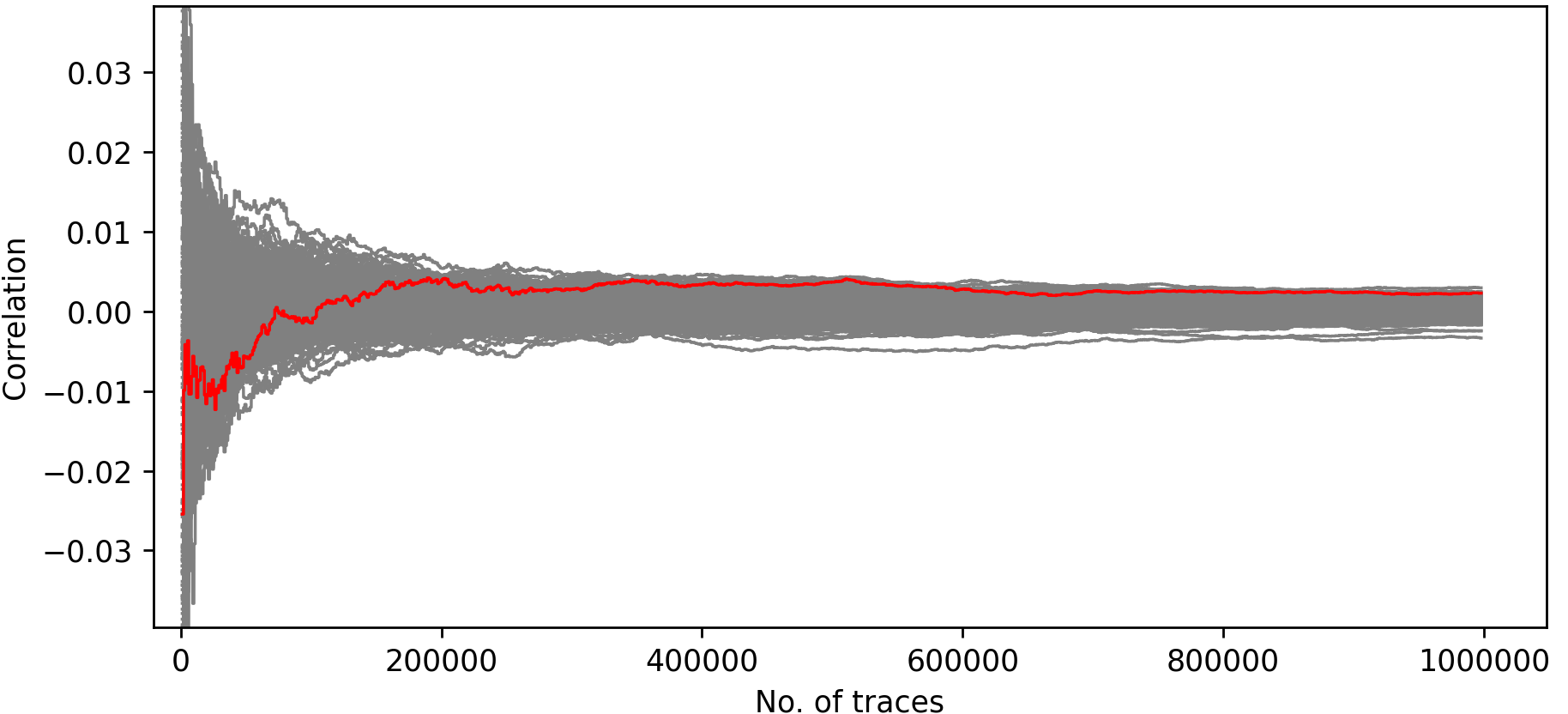}
    \caption{CPA over 1 million traces on 10th key byte of PLL \bt-protected \texttt{AES}, no emergence of correct candidate.}
    \label{fig:byte10_PLL}
\end{figure}

\begin{figure}[h]
    \centering
    \includegraphics[width=\linewidth]{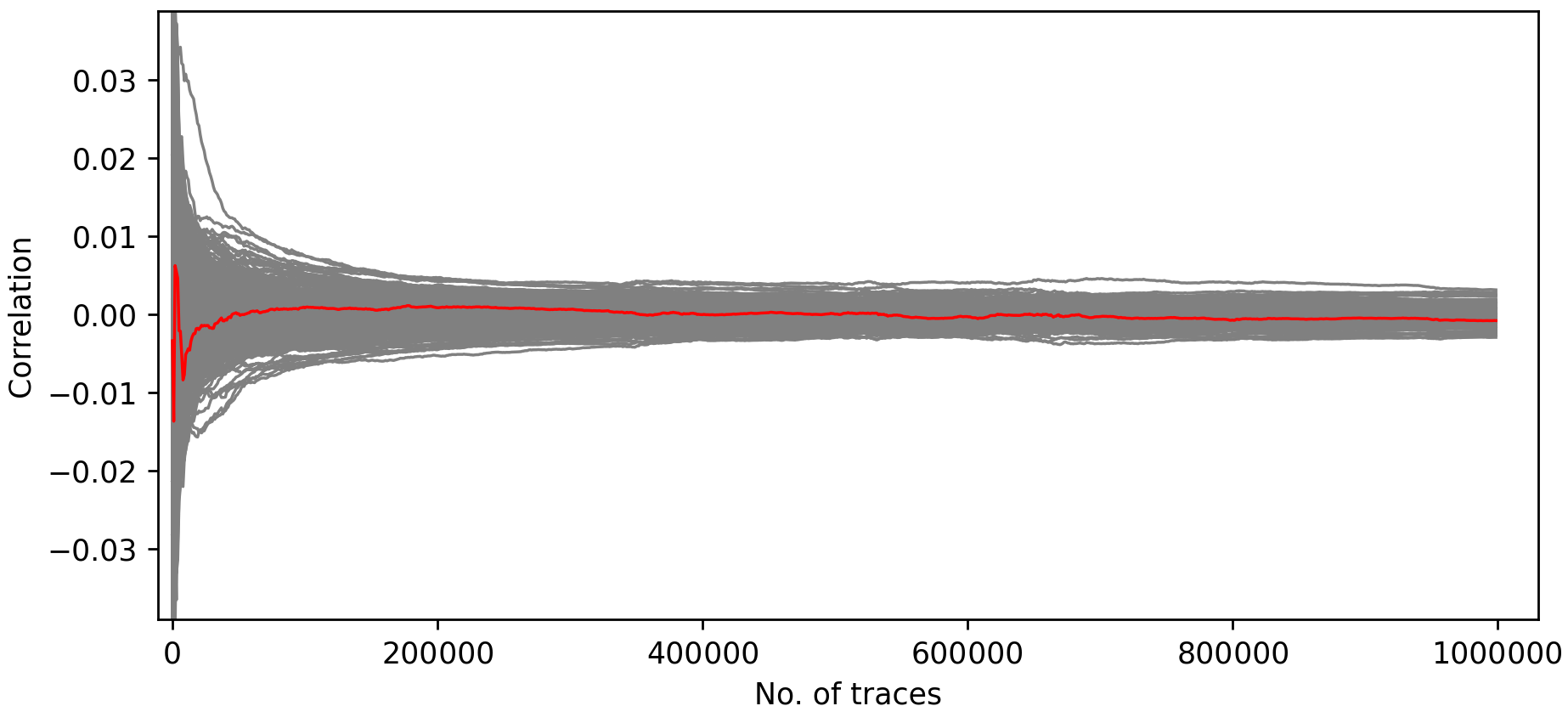} 
    \caption{Second-order CPA over 1 million traces on 2nd key byte of asynchronous \bt-protected first-order masked \texttt{SKINNY}, no emergence of correct candidate.}
    \label{fig:MTD_skinny_BT}
\end{figure}

In \cref{fig:byte10_PLL} we show an unsuccessful attack against our PLL-\bt-protected \texttt{AES} circuit.
Similarly, \cref{fig:MTD_skinny_BT} shows an unsuccessful attack against the \bt-protected \texttt{SKINNY} circuit.

\section{CPA Against \texttt{AES}}\label{app:cpaaes}

Here we describe the procedure for carrying out final-round Correlation Power Analysis (CPA) against \texttt{AES}~\cite{Rijndael, AESspec} to recover the secret key from side-channel leakage.

\parhead{Power Model.}
The first step for performing CPA is to select a power model that approximates the dependency of the power consumption of a CMOS circuit on the data being processed within.
The most commonly used leakage models are Hamming weight (HW) and Hamming distance (HD) (or `switching distance').
In the HW model~\cite{KocherJJ99}, power consumption is dependent on the number of bits that are set on (to 1) in stored data.
In the HD model~\cite{Brier04}, the consumption depends on how many bit transitions (0 $\rightarrow$ 1 or 1 $\rightarrow$ 0) occur in a computation.
Since we are concerned with static leakages which depend on fixed register contents, we use the HW power model against the stored intermediate, i.e.\ we expect that the power consumption for a given state is correlated with the number of register bits that are set to 1.

\parhead{Target Intermediate.}
Next we identify a computational intermediate (a value registered at some point in execution) that depends on a combination of known values (inputs and/or outputs) 
with a part of the secret value that has a relatively small guess space.
We opt for a final-round attack on \texttt{AES128} where the target intermediate \bm{$x_i$} is the $i^{th}$ byte of the output state from the penultimate (ninth) round, and final round SBox input. 
This intermediate depends on a combination of an adversary-known value (output ciphertext) and part of the secret key value with a small guess space. 
All final round \texttt{AES} operations are byte-wise since \mbox{\emph{MixColumns}} is not performed.
Reversing the final round, each byte of the intermediate can be derived using \Cref{eq:1} from the known corresponding output ciphertext byte \bm{$c_j$}, and the corresponding round subkey \bm{$k_j$} which has a 256-candidate guess space.
\begin{equation}
\small
\label{eq:1}
\bm{x_i} = \operatorname{{InverseSubBytes}}(\bm{c_j} \oplus \bm{k_j}) 
\end{equation}
$i$ is a given index of a byte of the ninth round output, and $i$ maps to $j$ after the final round \mbox{\emph{ShiftRows}} operation.

\parhead{Attack Procedure.}
We now describe last-round CPA against \texttt{AES}.
In the acquisition phase we perform many ($N$) encryptions with various random plaintexts, and for each we measure the power (trace) consumed by the device during a clock cycle when we know it stores the intermediate.
For a given encryption, the intermediate value should influence the power trace according to our hypothesised HW power model.
For each encryption we store a tuple of the power trace $P$ and the corresponding output ciphertext \bm{$c_j$}.

\begin{algorithm} 
\footnotesize
\begin{algorithmic}[1]
\caption{CPA against final round of \texttt{AES}}
\label{alg:loop}
\For{$i \gets 0$ to $15$} \Comment{For all key block bytes}
    \State $j \gets \operatorname{ShiftRowIndex}(i)$
    \For{$\mathit{traces} \gets 0$ to $N$}
        \For{$\bm{k_j} \gets 0$ to $255$} \Comment{For all subkey guesses}
            \State $\bm{x_i} \gets \operatorname{InverseSubBytes}(\bm{c_j} \oplus \bm{k_j})$ \Comment{Target intermediate}
            \State $\bm{H[k_j]}$.append($\operatorname{HW}(\bm{x_i})$) \Comment{$\bm{H}$: $256 \times \mathit{traces}$}
        \EndFor 
        \State $\bm{C}[\bm{k_j}] \gets correlation(\bm{H[k_j]}, \bm{P})$
        \State $\operatorname{Subkey} \operatorname{guess} \gets \operatorname{argmax}(|\bm{C}|)$
    \EndFor
\EndFor
\end{algorithmic}
\end{algorithm}

Separate from acquisition is the processing phase, outlined in \Cref{alg:loop}.
For this explanation we consider one subkey byte, i.e.\ one $i$ in \Cref{alg:loop}.
One trace at a time, for each of the 256 possible round subkey guesses we derive the value that the intermediate would assume based on the stored output and the key guess using \Cref{eq:1}, and we store the HW of this calculated intermediate (in $\bm{H}$).
We end up with a $256 \times N$ matrix where each row represents the expected power consumption for all traces for a given subkey guess.
The correlation between the expected power consumption for each subkey guess across all traces $\bm{H[k_j]}$ with the measured power consumption $\bm{P}$ is then calculated.
Given enough traces are gathered, only the correct subkey expected power consumption will exhibit correlation with the measured consumption.

For explanation's sake we distinguished acquisition and processing phases, describing the overall process for a limited number ($N$) of traces.
However, we actually perform these phases simultaneously rather than sequentially.
Moreover, we do not set out with a predefined number of traces ($N$), instead as we acquire more traces we repeatedly evaluate to see if a standout candidate has emerged.

\section{Second-Order CPA Against Masked \texttt{SKINNY}}\label{app:cpaskinny}
Here we describe the procedure for carrying out second-order Correlation Power Analysis (CPA) against \texttt{SKINNY}~\cite{Beierle16} to recover secret subkeys from side-channel leakage. 

\parhead{Target Intermediate.}
For our purposes we consider \texttt{SKINNY-128-128} with no tweak, i.e.\ all 128 tweakey bits are key material and therefore secret.
40 rounds are performed in this variant.
\texttt{SKINNY} has no key addition in either its first or last rounds.
Only half of the state (and key) is involved with key addition in each round, therefore retrieving the entire key requires attacking two encryption rounds, which depends on which subkey the attacker wants to retrieve. 
Our target intermediates are similarly the round output states which are inputs to subsequent round SBoxes.  

\parhead{Attack Procedure.}
The procedure for attacking the masked \texttt{SKINNY} implementation is mostly similar to that we previously describe for attacking our \texttt{AES} target, with a few differences.
We follow the same steps described in \cref{alg:loop}, but replacing the target intermediate (line 5) and \cref{eq:1} with the \texttt{SKINNY} intermediates.
Since we target a masked implementation we must perform a higher (second) order attack to leak the sensitive value from measurements of the masked intermediates. 
To do so, we subtract the average value across traces P from all traces then square the results.
This is before computing the correlation between traces P and intermediates for each subkey guess (line 8).

\end{document}